\documentstyle[12pt]{article}

\def\ba{\begin{eqnarray}}
\def\ea{\end{eqnarray}}

\def\be{\begin{equation}}
\def\ee{\end{equation}}

\begin{document}

\begin{center}
{\LARGE {Toric $G_{2}$ and $Spin(7)$ holonomy spaces from gravitational}}

{\LARGE {instantons and other examples}}

\smallskip 
\[
\]

{Gast\'on E. Giribet and Osvaldo P. Santill\'an}

\smallskip

{Physics Department, Universidad de Buenos Aires}

{\it Ciudad Universitaria, Pabell\'on I, 1428, Buenos Aires, Argentina.}

\smallskip \smallskip 
\[
\]

{\bf Abstract}
\end{center}

Non-compact $G_{2}$ holonomy metrics that arise from a $T^{2}$ bundle over a
hyper-K\"{a}hler space are constructed. These are one parameter deformations
of certain metrics studied by Gibbons, L\"{u}, Pope and Stelle in \cite%
{Stele}. Seven-dimensional spaces with $G_2$ holonomy fibered over the
Taub-Nut and the Eguchi-Hanson gravitational instantons are found, together
with other examples. By using the Apostolov-Salamon theorem \cite{Apostol},
we construct a new example that, still being a $T^2$ bundle over
hyper-K\"ahler, represents a non trivial two parameter deformation of the
metrics studied in \cite{Stele}. We then review the $Spin(7)$ metrics
arising from a $T^3$ bundle over a hyper-K\"ahler and we find two parameter
deformation of such spaces as well. We show that if the hyper-K\"ahler base
satisfies certain properties, a non trivial three parameter deformations is
also possible. The relation between these spaces with half-flat and almost $%
G_2$ holonomy structures is briefly discussed.

\bigskip

\section{Introduction}

The spaces of special holonomy have received remarkable attention within the
context of high energy physics; being the spaces of $G_{2}$ holonomy one of
the most important examples. This is due to the fact that these geometries,
in the presence of singularities, give rise to a natural framework for
reducing eleven-dimensional M-theory to ${\cal N}=1$ four-dimensional
realistic models \cite{AtiyahWitten,Witten,AcharyaWitten}; see also \cite%
{Gubserrrr, Jose}. Actually, it is a well known fact that the geometries of
this sort admit at least one globally defined covariantly constant spinor 
\cite{Bryant} and, when singularities are present, non-abelian gauge fields
and chiral matter can also appear. All these features make evident that the
study of spaces of $G_{2}$ holonomy deserved the attention of theoretical
physicists. However, this is actually a hard tool to be studied, mainly
because the explicit examples of compact spaces of $G_{2}$ holonomy have so
far eluded discovery. In fact, even though such spaces are known to exist 
\cite{Joyce}, the metrics which are explicitly known \cite{Bryant}-\cite{Lu}
turn out to be non-compact. Nevertheless, it was shown that the M-theory
dynamics near the singularity is not strongly dependent on the global
properties of the space \cite{AtiyahWitten}. On the other hand, applications
for which the presence of singularities plays no role also exist.
Furthermore, still in the case of non-compact spaces, it turns out that
finding and classifying metrics of holonomy $G_{2}$ represents a highly
non-trivial problem. In this work we discuss non-compact examples of toric
geometries of such special holonomy, which are based on four-dimensional
hyper-K\"{a}hler gravitational instantons.

The problem of classifying spaces of $G_{2}$ holonomy possessing one
isometry whose Killing vector orbits form a K\"{a}hler six-dimensional space
was analyzed both by physicist and mathematicians. In Ref. \cite{Kaste}, it
was concluded that such geometries are described by a sort of holomorphic
monopole equation together with a condition related to the integrability of
the complex structure. Such condition turns out to be stronger than the one
required by supersymmetry. On the other hand, Apostolov and Salamon have
proven in \cite{Apostol} that the K\"{a}hler condition yields the existence
of a new Killing vector that commutes with the first, so that these metrics
are toric. Besides, it was shown that such a $G_{2}$ metric yields a
four-dimensional manifold equipped with a complex symplectic structure and a
one-parameter family of functions and 2-forms linked by second order
equations (henceforth called Apostolov-Salamon equation). The inverse
problem, i.e. the one of constructing a torsion-free $G_{2}$ structure
starting from such a four-dimensional space was also discussed in \cite%
{Apostol}. Then, a natural question arises as to whether both description of
this classification problem are equivalent. In Ref \cite{yo}, it was argued
that this is indeed the case. Moreover, in Ref. \cite{Stele,Apostol,yo} such
a construction was employed to generate new $G_{2}$-metrics. In the present
work, the solution generating technique will be analyzed in detail and a
wider family of $G_{2}$-metrics will be written down. In particular, the
Eguchi-Hanson and the Taub-Nut metrics will be dimensionally extended to new
examples of $G_{2}$ holonomy following the construction proposed in \cite%
{Stele}. In section 2 we discuss the Apostolov-Salamon theorem \cite{Apostol}%
, which formalizes a method for systematically constructing spaces with
special holonomy $G_{2}$ by starting with a hyper-K\"{a}hler space in
four-dimensions. This construction is actually the one previously employed
by Gibbons, L\"{u}, Pope and Stelle in Ref. \cite{Stele} and here we discuss
it within the framework of \cite{Apostol}. Then, we describe some explicit
examples in order to illustrate the procedure. In particular, we show how
some of the $G_{2}$ metrics obtained in such way are one-parameter
deformations of those examples discussed in the literature. In section 3, we
present the examples that are based on non-trivial Gibbons-Hawking
solutions. We discuss the $G_{2}$ spaces obtained by starting with the
four-dimensional gravitational instantons and we write down the
corresponding metrics explicitly. Other examples are also discussed. In
particular, we find a two parameter deformations of the $T^{3}$ bundles with 
$Spin(7)$ holonomy over hyper-K\"{a}hler studied in \cite{Stele}. Besides,
in the strictly almost K\"{a}hler case, we obtain $Spin(7)$ metrics which
correspond to non trivial $T^{3}$ bundles over hyper-K\"{a}hler metrics. To
our knowledge, such metrics were not considered before in the literature. We
show that all the presented metric spaces are foliated by equidistant
hypersurfaces. In the $G_{2}$ holonomy case, these surfaces are half-flat
manifolds, while for the $Spin(7)$ metrics they are almost $G_{2}$ holonomy
spaces. So that we implicitly find a family of half-flat $T^{2}$ bundles and
a family of almost $G_{2}$ holonomy $T^{3}$ bundles over hyper-K\"{a}hler
spaces.

\section{The general setup}

Let us begin by explaining how the Apostolov-Salamon scheme works; detailed
proofs can be found in \cite{Apostol}. Let us consider a four dimensional
complex manifold $M$ with a metric $g_{4}=\delta _{ab}e^{a}\otimes e^{b}$.
It is also assumed that this metric is a function of certain parameter $\mu $%
, i.e. $g_{4}=g_{4}(\mu )$. This parameter should not be confused with a
coordinate of $g_{4}$. The property of $M$ as being \textquotedblleft
complex\textquotedblright\ means the following: Consider the manifold $M$,
for which there exists a $(1,1)$-tensor $J_{1}$ with the property $%
J_{1}\cdot J_{1}=-I$, such that $g_{4}(J_{1}\cdot ,\cdot )=-g_{4}(\cdot
,J_{1}\cdot )$. A metric for which the last property holds is called
hermitian with respect to $J_{1},$ and it follows that $\widetilde{J}%
_{1}=g_{4}(J_{1}\cdot ,\cdot )$ is an antisymmetric tensor, so that it is a $%
\mu $-dependent 2-form. In the cases in which there is no dependence on $\mu 
$ we will denote this tensor as $\overline{J}_{1}$. The tensor $J_{1}$ is
called an almost complex structure and is not uniquely defined. Indeed, any $%
SO(4)$ rotation of the frame $e^{a}$ induces a new almost complex structure
for which the metric turns out to be again hermitian. If at least one element
of such family of almost complex structures is covariantly constant with
respect to the Levi-Civita connection $\Gamma $ of $g_{4}$ (that is, $\nabla
_{X}J_{1}=0,$ $X$ being an arbitrary element of $TM$) then the tensor $J_{1}$
will be called a complex structure, and then the manifold $M$ will be
complex. Conversely, for any complex manifold it is possible to find at
least one complex structure. The condition $\nabla _{X}J_{1}=0$ is
equivalent to the condition $d\overline{J}_{1}=0$ (or $d_{M}\widetilde{J}%
_{1}=0$ if there is a dependence on $\mu $, being $d_{M}$ the
differentiation over $M$ which does not involve derivatives with respect to $%
\mu $) together with the integrability of $J_{1}$; that is, the vanishing of
the Nijenhuis tensor associated to $J_{1}$.

It will be assumed also that the metric $g_{4}$ admits a complex symplectic
form $\Omega =\overline{J}_{2}+i\overline{J}_{3}$, where being
``symplectic'' means that it is closed, $d\Omega =0$. On the other hand,
being ``complex'' implies that 
\[
\overline{J}_{2}\wedge \overline{J}_{2}=\overline{J}_{3}\wedge \overline{J}%
_{3},\qquad \overline{J}_{2}\wedge \overline{J}_{3}=0, 
\]%
and that 
\begin{equation}
\overline{J}_{2}(J_{1}\cdot ,\cdot )=\overline{J}_{3}(\cdot ,\cdot ).
\label{areva}
\end{equation}%
Now let us introduce a function $u$ depending on the coordinates of $M$ and
on the parameter $\mu $, and satisfying 
\begin{equation}
2\mu \widetilde{J}_{1}(\mu )\wedge \widetilde{J}_{1}(\mu )=u\Omega \wedge 
\overline{\Omega }.  \label{compota}
\end{equation}%
This function always exists because the wedge products appearing in (\ref%
{compota}) are proportional to the volume form of $M$. With the help of the
quantities defined above a seven-dimensional metric \cite{Apostol,Stele} 
\begin{equation}
g_{7}=\frac{(d\alpha +H_{2})^{2}}{\mu ^{2}}+\mu \;\left( \;u\;d\mu ^{2}+%
\frac{(d\beta +H_{1})^{2}}{u}+g_{4}(\mu )\;\right) ,  \label{senio}
\end{equation}%
can be constructed. Here $\beta $ and $\alpha $ are two new coordinates
while $H_{1}$ and $H_{2}$ are certain 1-forms independent on $\beta $ and $%
\alpha $. The parameter $\mu $ of $g_{4}$ is now a coordinate of $g_{7}$ and
the metric tensor $g_{7}$ is also independent on the coordinates $\alpha $
and $\beta $. This means that the vectors $\partial _{\alpha }$ and $%
\partial _{\beta }$ are obviously Killing and commuting and therefore the
metrics (\ref{senio}) are to be called ``toric''. The Apostolov-Salamon
construction states that if the quantities appearing in (\ref{senio}) are
related by the evolution equation 
\begin{equation}
\frac{\partial ^{2}\widetilde{J}_{1}}{\partial ^{2}\mu }=-d_{M}d_{M}^{c}u,
\label{ebol}
\end{equation}%
and the forms $H_{1}$ and $H_{2}$ are defined on $M\times {\bf R}_{\mu }$
and $M$ respectively by the equations 
\begin{equation}
dH_{1}=(d_{M}^{c}u)\wedge d\mu +\frac{\partial \widetilde{J}_{1}}{\partial
\mu },\qquad dH_{2}=-\overline{J}_{2},  \label{chon}
\end{equation}%
with $d_{M}^{c}=J_{1}d_{M}$, then the seven-dimensional metric (\ref{senio})
has holonomy in $G_{2}$. Moreover, the six-dimensional metric%
\begin{equation}
g_{6}=u\;d\mu ^{2}+\frac{(d\beta +H_{1})^{2}}{u}+g_{4}(\mu ),  \label{rel}
\end{equation}%
is K\"{a}hler with K\"{a}hler form 
\begin{equation}
K=(d\beta +H_{1})\wedge d\mu +\widetilde{J}_{1}.  \label{chu}
\end{equation}%
This condition is usually referred as the ``strong supersymmetry condition''
in the physical literature \cite{Kaste}. The calibration $3$-form
corresponding to the metrics (\ref{senio}) is 
\[
\Phi =\widetilde{J}_{1}(\mu )\wedge (d\alpha +H_{2})+d\mu \wedge (d\beta
+H_{1})\wedge (d\alpha +H_{2}) 
\]%
\begin{equation}
+\mu \;\left( \;\overline{J}_{2}\wedge (d\beta +H_{1})+u\overline{J}%
_{3}\wedge d\mu \;\right) ,  \label{dale}
\end{equation}%
and by means of (\ref{chon}), (\ref{ebol}) and (\ref{compota}) it can be
seen directly seen that $d\Phi =d\ast \Phi =0$. This is an standard feature
of the reduction of the holonomy from $SO(7)$ to $G_{2}$ \cite{Bryant}.
Moreover, the Killing field $\partial _{\alpha }$ preserves $\Phi $ and $%
\ast \Phi $.

The converse of all these statements are also true. That is, for given a $%
G_{2}$ holonomy manifold $Y$ with a metric $g_{7}$ possessing a Killing
vector that preserves the calibration forms $\Phi $ and $\ast \Phi $ and
such that the six-dimensional metric $g_{6}$ obtained from the orbits of the
Killing vector is K\"{a}hler, then there exists a coordinate system in which 
$g_{7}$ takes the form (\ref{senio}) being $g_{4}(\mu )$ a one-parameter
four-dimensional metric admitting a complex symplectic structure $\Omega $
and a complex structure $J_{1}$, being the quantities appearing in this
expression related by (\ref{areva}) and the conditions (\ref{chon}), (\ref%
{ebol}) and (\ref{compota}). Details of these assertions can be found in the
original reference \cite{Apostol}.

\subsection{ $G_2$ holonomy metrics fibered over hyper-K\"ahler manifolds}

It is worth mentioning that the integrability conditions 
\[
d\bigg((d_{M}^{c}u)\wedge d\mu +\frac{\partial \widetilde{J}_{1}}{\partial
\mu }\bigg)=0,\qquad d\overline{J}_{2}=0 
\]%
for (\ref{chon}) are identically satisfied. The second is the closure of $%
\overline{J}_{2}$ while the first is a direct consequence of (\ref{ebol})
together with the closure of $\widetilde{J}_{1}$. Equations (\ref{chon}) are
considerably simplified when $d_{M}^{c}u=0$. In this case, the first
equation in (\ref{chon}) and the condition $d^{2}H_{1}=0$ imply that $%
\partial \widetilde{J}_{1}/\partial \mu $ should be $\mu $-independent and
closed. If this is so then (\ref{ebol}) is trivially satisfied. One possible
solution is to choose 
\[
\widetilde{J}_{1}(\mu )=(M+Q\mu )\overline{J}_{1}, 
\]%
being $\overline{J}_{1}$ $\mu $-independent and closed, and where $M$ and $Q$
certain real parameters. The metric $g_{4}$ for which $\widetilde{J}_{1}$ is
a K\"{a}hler form is 
\begin{equation}
g_{4}(\mu )=(M+Q\mu )\widetilde{g}_{4},  \label{o}
\end{equation}%
being $\widetilde{g}_{4}$ a $\mu $-independent metric and $\overline{J}_{1}$
its K\"{a}hler form. Relation (\ref{compota}) gives a simple algebraic
equation for $u$, yielding 
\[
u=\mu (M+Q\mu )^{2}. 
\]%
In conclusion, the metric $\widetilde{g}_{4}$ is quaternion hermitian with
respect to the tensors $J_{i},$ and from (\ref{areva}) it follows that $%
J_{i}\cdot J_{j}=-\delta _{ij}+\epsilon _{ijk}J_{k}$. Besides, the two forms 
$\overline{J}_{1}$, $\overline{J}_{2}$ and $\overline{J}_{3}$ are closed. In
four dimensions the closure of the hyper-K\"{a}hler triplet implies that the 
$J_{i}$ are integrable. This means that $\widetilde{g}_{4}$ is hyper-K\"{a}%
hler. The $G_{2}$ holonomy metric corresponding to this case is then given
by 
\begin{equation}
g_{7}=\frac{(d\beta +QH_{1})^{2}}{(M+Q\mu )^{2}}+\frac{(d\alpha +H_{2})^{2}}{%
\mu ^{2}}+\mu ^{2}\;(M+Q\mu )^{2}\;d\mu ^{2}+\mu \;(M+Q\mu )\;\widetilde{g}%
_{4},  \label{metricol}
\end{equation}%
and the 1-forms $H_{1}$ and $H_{2}$ obey $dH_{1}=\overline{J}_{1}$ and $%
dH_{2}=-\overline{J}_{2},$ being $\overline{J}_{1}$ and $\overline{J}_{2}$
any pair of the three K\"{a}hler forms on $M$. It is important to remark
that (\ref{metricol}) gives three $G_{2}$ holonomy metrics for a given
hyper-K\"{a}hler metric. This is because one pair of 2-forms, selected among
those that form the hyper-K\"{a}hler triplet $\overline{J}_{i}^{\ast },$ is
necessary. I.e. although there are six possible choices, the order of the
selection has no relevance and there are essentially three pairs. If the
parameter $M$ is set to zero, then the resulting metrics correspond to those
appearing in section 6.2 of reference \cite{Stele}. Although (\ref{metricol}%
) are just a subfamily of the Apostolov-Salamon metrics, the extension (\ref%
{metricol}) gives rise to a powerful method to construct new $G_{2}$
examples. In principle, these have to be distinguished from other $G_{2}$
metrics presented in the literature \cite{Santillan}-\cite{Lust}, which are
of the Bryant-Salamon type \cite{Bryant}. Regarding this, let us briefly
comment on the amount of effective parameters appearing in both
constructions. For Apostolov-Salamon metrics (\ref{metricol}) there appear
two parameters $M$ and $Q$ but only one of them is an effective one. It is
not difficult to see that if $M\neq 0$ then it can be set $1$ by simply
rescaling as%
\[
\widetilde{g}_{4}\rightarrow M\widetilde{g}_{4},\qquad H_{1}\rightarrow
MH_{1},\qquad H_{2}\rightarrow MH_{2} 
\]%
\[
\alpha \rightarrow M\alpha ,\qquad \beta \rightarrow M^{2}\beta ,\qquad
Q\rightarrow Q^{\prime }=\frac{Q}{M} 
\]%
Then, the number of effective parameters will be $1+n$ being $n$ the numbers
of those belonging to the base 4-space. On the other hand, if $M=0$, then we
can also make the following redefinitions 
\[
\widetilde{g}_{4}\rightarrow Q\widetilde{g}_{4},\qquad H_{1}\rightarrow
QH_{1},\qquad H_{2}\rightarrow QH_{2} 
\]%
\[
\alpha \rightarrow Q\alpha ,\qquad \beta \rightarrow Q^{2}\beta , 
\]%
and set $Q=1$. In this case, the only parameters appearing in the metric
will be those $n$ of the hyper-K\"{a}hler base.

There is another way to check that $g_{7}$ is given by (\ref{metricol}). Let
us define the tetrad 1-forms 
\begin{equation}
e^{5}=\mu \;(M+Q\mu )\;d\mu ,\qquad e^{6}=\frac{d\alpha +H_{2}}{\mu },\qquad
e^{7}=\frac{d\beta +QH_{1}}{M+Q\mu }.  \label{pip}
\end{equation}%
Then calibration form (\ref{dale}) is then expressed as 
\begin{equation}
\Phi =\mu \;(M+Q\mu )\;\overline{J}_{1}\wedge e^{6}+e^{5}\wedge e^{6}\wedge
e^{7}+\mu \;(M+Q\mu )\;\left( \;\overline{J}_{2}\wedge e^{7}+\overline{J}%
_{3}\wedge e^{5}\;\right) .  \label{dale2}
\end{equation}%
It is convenient to choose a tetrad basis $\widetilde{e}^{i}$ for which the
hyper-K\"{a}hler metric is diagonal, i.e. $g_{4}=\delta _{ab}\widetilde{e}%
^{a}\otimes \widetilde{e}^{b},$ and for which the hyper-K\"{a}hler triplet
takes the form 
\begin{equation}
\overline{J}_{1}=\widetilde{e}^{1}\wedge \widetilde{e}^{2}+\widetilde{e}%
^{3}\wedge \widetilde{e}^{4},\qquad \overline{J}_{2}=\widetilde{e}^{1}\wedge 
\widetilde{e}^{3}+\widetilde{e}^{4}\wedge \widetilde{e}^{2},\qquad \overline{%
J}_{3}=\widetilde{e}^{1}\wedge \widetilde{e}^{4}+\widetilde{e}^{2}\wedge 
\widetilde{e}^{3}.  \label{posoto}
\end{equation}%
Then, by making the redefinition 
\[
e^{a}=\mu ^{1/2}\;(M+Q\mu )^{1/2}\widetilde{e}^{a} 
\]%
we see that (\ref{dale2}) takes the familiar octonionic form 
\[
\Phi =c_{abc} \ e^{a}\wedge e^{b}\wedge e^{c}, 
\]%
where $c_{abc}$ is the octonion constants. The form $\Phi $ is $G_{2}$
invariant as a consequence of the fact that $G_{2}$ is the automorphism
group of the octonion algebra. The $G_{2}$ holonomy metric corresponding to $%
\Phi $ is simply $g_{7}=\delta _{ab}e^{a}\otimes e^{b}$ and it can be
checked that $g_{7}$ is indeed given by (\ref{metricol}). This follows from
the expressions for $e^{a}$ given above. It is convenient to remark that, in
(\ref{posoto}), we were assuming that the hyper-K\"{a}hler triplet is
positive oriented. In the negative oriented case it can be analogously shown
that the holonomy will be also $G_{2}$. \newline

\subsection{The simplest example}

An interesting example to illustrate this construction comes from
considering the simplest hyper-K\"{a}hler 4-manifold, namely ${\bf R}^{4}$
provided with its flat metric ${\bf E}^{4}=g_{4}=dx^{2}+dy^{2}+dz^{2}+dt^{2}$%
. A closed hyper-K\"{a}hler triplet for ${\bf R}^{4}$ is given by%
\begin{equation}
\overline{J}_{1}=dt\wedge dy-dz\wedge dx,\qquad \overline{J}_{2}=dt\wedge
dx-dy\wedge dz,\qquad \overline{J}_{3}=dt\wedge dz-dx\wedge dy.  \label{pot}
\end{equation}%
Now, let us extend this example to a seven-dimensional metric by means of (%
\ref{metricol}). This innocent looking case is indeed rather rich and
instructive. As it will be explained below, it gives rise to metric with
holonomy exactly $G_{2}$. This example was already discussed in Ref. \cite%
{Santillan} and we will extend it here to less simple cases.

As it has been mentioned, there are three possible $G_{2}$ metrics that can
be constructed, depending on which pair of forms we select from (\ref{pot}).
But in the flat case, this choice just corresponds to a permutation of
coordinates and the resulting metrics will be actually the same. By
selecting the first two $\overline{J}_{i}$ among those in (\ref{pot}) we
obtain the potential forms%
\[
H_{1}=-xdz-ydt, 
\]%
\[
H_{2}=-ydz-xdt. 
\]%
Hence, the corresponding $G_{2}$ holonomy metrics read%
\begin{equation}
g_{7}=\frac{(\;d\beta -Q(xdz+ydt)\;)^{2}}{(M+Q\mu )^{2}}+\frac{(d\alpha
-ydz-xdt)^{2}}{\mu ^{2}}+\mu ^{2}\;(M+Q\mu )^{2}\;d\mu ^{2}  \label{plani}
\end{equation}%
\[
+\mu \;(M+Q\mu )(\;dx^{2}+dy^{2}+dz^{2}+dt^{2}\;). 
\]%
If we select $M=0$ and $Q=1$ the metric tensors (\ref{plani}) reduces to 
\begin{equation}
g_{7}=\frac{(d\beta -xdz+ydt)^{2}}{\mu ^{2}}+\frac{(d\alpha -ydz-xdt)^{2}}{%
\mu ^{2}}+\mu ^{4}\;d\mu ^{2}+\mu ^{2}\;(\;dx^{2}+dy^{2}+dz^{2}+dt^{2}\;).
\label{plani2}
\end{equation}%
Actually, metrics (\ref{plani2}) have been already obtained in the
literature \cite{Stele}. They have been constructed within the context of
eleven-dimensional supergravity, by starting with a domain wall solution of
the form 
\[
g_{5}=H^{4/3}(\;dx^{2}+dy^{2}+dz^{2}+dt^{2}\;)+H^{16/3}d\mu ^{2}, 
\]%
being $a=1,..,4$ and $H$ a warp function (see \cite{Stele} for the details).
By making use of the often called oxidation rules, and by starting from the
above space in five-dimensions, it is feasible to get a eleven-dimensional
background of the form $g_{11}=g_{(3,1)}+g_{7}$ with the seven-dimensional
metric being%
\begin{equation}
g_{7}=\frac{(d\beta -xdz+ydt)^{2}}{H^{2}}+\frac{(d\alpha -ydz-xdt)^{2}}{H^{2}%
}+H^{4}\;d\mu ^{2}+H^{2}\;(\;dx^{2}+dy^{2}+dz^{2}+dt^{2}\;).  \label{oxo}
\end{equation}%
By selecting the homothetic case $H=\mu $ the last metric reduces to (\ref%
{plani2}). It can be shown by explicit calculation of the curvature tensor
that (\ref{oxo}) is irreducible and has holonomy exactly $G_{2}$, and this
happens even in the particular case $H=\mu ,$ i.e. the one in (\ref{plani2}%
), cf. \cite{Stele}. The isometry corresponding to (\ref{oxo}) and (\ref%
{plani}) is the same $SU(2)$ group that acts linearly on the coordinates $%
(x,y,z,t)$ on $M$ and which simultaneously preserves the two forms $%
\overline{J}_{1}$ and $\overline{J}_{2}$. The $SU(2)$ action on $M$ supplies 
$H_{1}$ and $H_{2}$ with total differential terms that can be absorbed by a
redefinition of the coordinates $\alpha $ and $\beta $. For instance, a
general translation of the form 
\begin{equation}
x\rightarrow x+\alpha _{1},\qquad y\rightarrow y+\alpha _{2},\qquad
z\rightarrow z+\alpha _{3},\qquad t\rightarrow t+\alpha _{4},  \label{sip}
\end{equation}%
does preserve $\overline{J}_{1}$ and $\overline{J}_{2}$ but does not
preserve the 1-forms $H_{1}$ and $H_{2}$. Nevertheless, the effect of the
translations (\ref{sip}) can be compensated by a coordinate transformation
of the form 
\begin{equation}
\beta \rightarrow \beta +\alpha _{5}+\alpha _{1}z-\alpha _{2}t,\qquad \alpha
\rightarrow \alpha +\alpha _{6}+\alpha _{2}z+\alpha _{1}t.  \label{sop}
\end{equation}%
Besides, more general $SU(2)$ transformations on $M$ preserving $\overline{J}%
_{1}$ and $\overline{J}_{2}$ can be also absorbed by a redefinition of the
coordinates $\alpha $ and $\beta $. For the metric (\ref{plani2}) we also
have the scale invariance under transformation 
\[
x\rightarrow \lambda x,\qquad y\rightarrow \lambda y,\qquad z\rightarrow
\lambda z,\qquad t\rightarrow \lambda t, 
\]%
\[
\alpha \rightarrow \lambda ^{4}\alpha ,\qquad \beta \rightarrow \lambda
^{4}\beta ,\qquad \mu \rightarrow \lambda ^{2}\mu , 
\]%
for a real parameter $\lambda $, which is generated by the homothetic
Killing vector 
\begin{equation}
D=2x\partial _{x}+2y\partial _{y}+2z\partial _{z}+2t\partial _{t}+\mu
\partial _{\mu }+4\alpha \partial _{\alpha }+4\beta \partial _{\beta }.
\label{homo}
\end{equation}%
Our task now is to construct more elaborated examples by means of similar
procedure. Let us move into this.

\section{$G_2$ metrics with three commuting Killing vectors}

By construction, metrics (\ref{metricol}) posses at least two Killing
vectors. If a larger isometry group is desired, an inspection of the formula
(\ref{metricol}) shows that the hyper-K\"{a}hler basis $\widetilde{g}_{4}$
should already posses Killing vectors and that the action of the isometry
group on $H_{1}$ and $H_{2}$ should be induced by gauge transforming $%
H_{1}\rightarrow H_{1}+df_{1}$ and $H_{2}\rightarrow H_{2}+df_{2}.$ The
effect of this transformation can be compensated by a redefinition of the
coordinates $\alpha \rightarrow \alpha +f_{1}$ and $\beta \rightarrow \beta
+f_{2},$ so that the local form of the metric would result unaltered.
Considering $dH_{1}\sim \overline{J}_{1}$ and $dH_{2}\sim \overline{J}_{2}$
we see that $\overline{J}_{2}$ and $\overline{J}_{3}$ will be actually
preserved by the isometry group. An obvious example is an hyperkahler metric
with a the Killing vector which is tri-holomorphic, namely, one satisfying 
\[
{\cal L}_{K}\overline{J}_{1}={\cal L}_{K}\overline{J}_{2}={\cal L}_{K}%
\overline{J}_{3}=0. 
\]%
For any metric admitting a tri-holomorphic Killing vector $\partial _{t}$
there exists a coordinate system in which it takes generically the
Gibbons-Hawking form \cite{Gibbhawk} 
\begin{equation}
g=V^{-1}(dt+A)^{2}+Vdx_{i}dx_{j}\delta ^{ij},  \label{ashgib}
\end{equation}%
with a 1-form $A$ and a function $V$ satisfying the linear system of
equations 
\begin{equation}
\nabla V=\nabla \times A.  \label{Gibb-Hawk}
\end{equation}%
These metrics are hyper-K\"{a}hler with respect to the hyper-K\"{a}hler
triplet 
\[
\overline{J}_{1}=(dt+A)\wedge dx-Vdy\wedge dz 
\]%
\begin{equation}
\overline{J}_{2}=(dt+A)\wedge dy-Vdz\wedge dx  \label{transurop3}
\end{equation}%
\[
\overline{J}_{3}=(dt+A)\wedge dz-Vdx\wedge dy 
\]%
which is actually $t$-independent. The isometry group of the total $G_{2}$
space will be then enlarged to $T^{3}$.

Actually, one could naively suggest another possibility; namely, to choose a
Killing vector which is not actually tri-holomorphic, but still preserves
two of the three K\"{a}hler forms. However, as it will be shown below, an
isometry preserving $\overline{J}_{1}$ and $\overline{J}_{2}$ is necessarily
tri-holomorphic. If, on the other hand, a hyper-K\"{a}hler metric possesses
an isometry that is not tri-holomorphic, then there always exists a
coordinate system $(x,y,z,t)$ for which the distance element takes the form 
\cite{Boyer} 
\begin{equation}
g_{h}=u_{z}\;\left( e^{u}(dx^{2}+dy^{2})+dz^{2}\right) +u_{z}^{-1}\;\;\left(
dt+(u_{x}dy-u_{y}dx)\right) ^{2},  \label{gegodas}
\end{equation}%
with $u$ a function of $(x,y,z)$ satisfying the $SU(\infty )$ Toda equation;
namely%
\begin{equation}
(e^{u})_{zz}+u_{yy}+u_{xx}=0;
\end{equation}%
where we are denoting $f_{x^{i}}=\partial _{x^{i}}f$. It is evident that the
vector field $\partial _{t}$ is a Killing vector of (\ref{gegodas}). Metric (%
\ref{gegodas}) is thus hyper-K\"{a}hler with respect to the $t$-dependent
hyper-K\"{a}hler triplet 
\begin{eqnarray}
\overline{J}_{1} &=&e^{u}\ u_{z}dx\wedge dy+dz\wedge \left(
dt+(u_{x}dy-u_{y}dx)\right) , \\
\overline{J}_{2} &=&e^{u/2}\ \cos (t/2)\widetilde{\overline{J}}_{2}+e^{u/2}\
\sin (t/2)\widetilde{\overline{J}}_{3}, \\
\overline{J}_{3} &=&e^{u/2}\ \sin (t/2)\widetilde{\overline{J}}_{2}-e^{u/2}\
\cos (t/2)\widetilde{\overline{J}}_{3};  \label{beatit}
\end{eqnarray}%
where we defined the 2-forms%
\[
\widetilde{\overline{J}}_{2}=-u_{z}dz\wedge dy+\left( dt+u_{y}dx\right)
\wedge dy,\qquad \widetilde{\overline{J}}_{3}=u_{z}dz\wedge dx+\left(
dt+u_{x}dy\right) \wedge dx. 
\]%
These should not be confused with the hyper-K\"{a}hler forms $\overline{J}%
_{i}$ in (\ref{beatit}). From (\ref{beatit}) it is clear that $\partial _{t}$
preserve $\overline{J}_{1}$, but the other two K\"{a}hler forms turns out to
be dependent on $t$. It means that in the non-tri-holomorphic case it is
impossible to preserve two of the three $\overline{J}_{i}$ without
preserving the third one. The local form (\ref{gegodas}) is general enough,
and therefore, in four dimensions, a $U(1)$ isometry that preserves two of
the closed K\"{a}hler forms of an hyper-K\"{a}hler metric is automatically
tri-holomorphic.\footnote{%
This discussion concerns only $U(1)^n$ isometry groups, in other case it
does not apply.}

Actually, there exists a shorter argument in order to find the same
conclusion. All the complex structures form a two-sphere, since $u_1 J_1+u_2
J_2+u_3 J_3$ is a complex structure so long as $u_1^2+u_2^2+u_3^2=1$. Now
any U(1) action on a two sphere either is trivial and keeps all the complex
structures inert or keeps only two points (i.e. only one complex structure
and its conjugate) on the sphere fixed.\footnote{%
We thank S. Cherkis for this argument and for other important suggestions.}

\subsection{On $G_2$ metrics and asymptotic behavior of ALG spaces}

Let us consider now the Gibbons-Hawking metrics (\ref{ashgib}) for which the
potential $V$ and a vector potential $A$ are independent on certain
coordinate, say $x$. It is simple to check that the equations (\ref%
{Gibb-Hawk}) reduce, up to a gauge transformation, to the Cauchy-Riemann
equations 
\[
A_{y}=A_{z}=0,\qquad \partial _{y}A_{x}=-\partial _{z}V,\qquad \partial
_{z}A_{x}=\partial _{y}V. 
\]%
This means that we can write $A_{x}+iV=\tau _{1}+i\tau _{2}=\tau (w)$ with $%
w=z+iy$; namely, $A_{x}$ and $V$ turn out to be the real and the imaginary
part of an holomorphic function $\tau $ of one complex variable $w$. The
corresponding hyper-K\"{a}hler metric can be written as 
\begin{equation}
g_{4}=\frac{|dt+\tau dx|^{2}}{\tau _{2}}+\tau _{2}dwd\overline{w}.
\label{stew}
\end{equation}%
The \textquotedblleft local aspect\textquotedblright\ of the isometry group
corresponds to ${\bf R}^{2}$, ${\bf R}\times U(1)$ or $T^{2}=U(1)\times U(1)$%
, depending on the range of the values of the coordinates $x$ and $t$.
Notice that the Cauchy-Riemann equations imply that the 1-forms 
\begin{equation}
B_{1}=\tau _{1}dz-\tau _{2}dy,\qquad B_{2}=\tau _{1}dy+\tau _{2}dz
\label{trans}
\end{equation}%
are closed. Therefore, two functions $\varrho (x,y)$ and $\omega (x,y)$ such
that $d\varrho =B_{1}$ and $d\omega =B_{2}$ can be \textquotedblleft
locally\textquotedblright\ defined. It is easy to see that the $B_{i}$ are
the real and imaginary parts of the complex 1-form 
\[
B_{1}+iB_{2}=\tau (w)dw=dT(w), 
\]%
being $T(w)$ the primitive of the function $\tau $. Therefore $\varrho
+i\omega =T(w)$. With the help of this functions the hyper-K\"{a}hler
triplet for the family (\ref{stew}) is expressed as 
\[
\overline{J}_{1}^{\ast }=\Re \left( dt\wedge dx+\frac{1}{2}dw\wedge d%
\overline{T}\right) 
\]%
\begin{equation}
\overline{J}_{2}^{\ast }=\Im \left( \Upsilon \right) ,\qquad \overline{J}%
_{3}^{\ast }=\Re \left( \Upsilon \right) ,  \label{transurop6}
\end{equation}%
where we introduced the complex two 2-form 
\[
\Upsilon =dt\wedge dw+dx\wedge dT, 
\]%
and where the symbols $\Re $ and $\Im $ denote the real and the imaginary
part of the quantity between parenthesis. It is clear that both $\partial
_{t}$ and $\partial _{x}$ preserve (\ref{transurop6}). Therefore $\partial
_{x}$ is also a tri-holomorphic Killing vector, which clearly commutes with $%
\partial _{t}$. The 1-forms $H_{i}^{\ast }$ satisfying $dH_{i}^{\ast }=%
\overline{J}_{i}^{\ast }$ are easily found from (\ref{transurop6}), the
result is 
\[
H_{1}^{\ast }=-\Re \left( xdt+\frac{1}{2}\overline{T}dw\right) ,\qquad
H_{3}^{\ast }+iH_{2}^{\ast }=-wdt-Tdx, 
\]%
up to total differential terms. By selecting $H_{1}=H_{1}^{\ast }$ and $%
H_{2}=H_{2}^{\ast }$ in (\ref{metricol}) the following $G_{2}$ holonomy
metric is obtained 
\[
g_{7}=\frac{\;\left( d\beta -Q\;\Re (xdt+\frac{1}{2}\overline{T}dw)\;\right)
^{2}}{(1+Q\mu )^{2}}+\frac{\left( \;d\alpha -\Im (wdt+Tdx)\;\right) ^{2}}{%
\mu ^{2}}+\mu ^{2}\;(1+Q\mu )^{2}\;d\mu ^{2} 
\]%
\begin{equation}
+\mu \;(1+Q\mu )\;\left( \frac{|dt+\tau dx|^{2}}{\tau _{2}}+\tau _{2}dwd%
\overline{w}\;\right) ,  \label{metricoles}
\end{equation}%
Metrics (\ref{metricoles}) constitute a family of $G_{2}$ holonomy metrics
constructed essentially from a single holomorphic function $\tau $ and its
primitive $T$. There are two more $G_{2}$ holonomy metrics obtained by
selecting $H_{1}=H_{1}^{\ast }$ and $H_{2}=H_{3}^{\ast }$, and also $%
H_{1}=H_{2}^{\ast }$ and $H_{2}=H_{3}^{\ast }$ in (\ref{metricoles}). For
conciseness, we will not write them explicitly here, but the procedure to
find them follows straightforwardly. In all the cases there will be four
commuting Killing vectors namely, $\partial _{\alpha }$, $\partial _{\beta }$%
, $\partial _{t}$ and $\partial _{x}$.

Metrics of the form (\ref{stew}) have been considered in the physical
literature. For instance, the asymptotic form of any \textquotedblleft
ALG\textquotedblright\ instanton is (\ref{stew}) if the two coordinates $x$
and $t$ are periodically identified. Several examples of such geometries
have been considered in \cite{Cherkis,Gabo}. Let us recall that the term ALG
usually stands for a complete elliptically fibered hyper-K\"{a}hler
manifold. The metric (\ref{stew}) is not an ALG metric and in general is not
complete; however, it is what an ALG metric approaches at infinity. Another
context in which (\ref{stew}) have been considered in the context of stringy
cosmic strings \cite{Vafayau}. Besides, a particular case of such class of
metrics were shown to describe the single matter hypermultiplet target space
for type IIA superstrings compactified on a Calabi-Yau threefold when
supergravity and D-instanton effects are switched off \cite{Vafa}. In this
case we have $\tau =\log (w)$ with $T(w)=w(\log (w)-1)$, which, from (\ref%
{metricoles}), it yields the following explicit $G_{2}$ holonomy metric 
\[
g_{7}=\frac{\left( \;d\alpha -\Im \left( wdt+w(\log (w)-1\;)dx\right)
\right) ^{2}}{\mu ^{2}}+\frac{\left( \;d\beta -Q\;\Re (xdt+\frac{w}{2}%
(\;\log (w)-1)dw\;)\;\right) ^{2}}{(1+Q\mu )^{2}} 
\]%
\begin{equation}
+\mu ^{2}\;(1+Q\mu )^{2}\;d\mu ^{2}+\;\mu \;(1+Q\mu )\;\left( \frac{%
|\;dt+\log (w)dx\;|^{2}}{\log |w|}+\log |w|\;dwd\overline{w}\right) .
\label{metrice}
\end{equation}%
The base hyper-K\"{a}hler metric possesses a rotational Killing vector $%
\partial _{\arg (w)}$. However, such vectors do not preserve the hyper-K\"{a}%
hler triplet (\ref{transurop6}) and, for this reason, they will not be a
Killing vector of the seven-metric (\ref{metrice}).

It is interesting to analyze the transformation properties of the generic
metric (\ref{stew}) under the $SL(2,{\bf R})$ action 
\[
\tau \rightarrow \frac{a\tau +b}{c\tau +c},\qquad t\rightarrow at-bx,\qquad
x\rightarrow dx-ct, 
\]%
where the parameters $a$, $b$, $c$, $d$ satisfy $ad-bc=1$. Then, we obtain
that 
\[
\tau _{2}\rightarrow \frac{ad\tau _{2}}{|c\tau +d|^{2}},\qquad 
\]%
and the metric (\ref{stew}) transforms as 
\[
g_{4}=\frac{|dt+\tau dx|^{2}}{\tau _{2}}+\tau _{2}\frac{dwd\overline{w}}{%
|c\tau +d|^{2}}. 
\]%
By defining new complex coordinates $\xi ,$ by 
\[
d\xi =\frac{dw}{c\tau +d}, 
\]%
the metric becomes 
\[
g_{4}=\frac{|dt+\tau dx|^{2}}{\tau _{2}}+\tau _{2}d\xi d\overline{\xi }, 
\]%
which is of the form (\ref{stew}) and therefore hyper-K\"{a}hler. However,
notice the transformed metric is different than the former one, because $%
\tau $ will be a different function after the coordinate transformation $%
w\rightarrow \xi $. Therefore, in principle, the $SL(2,{\bf R})$ action is
not a symmetry of (\ref{stew}), but instead it maps any element of the
family of toric hyper-K\"{a}hler metric (\ref{stew}) into another one;
namely, it is closed among such family and can be regarded as a subgroup of
the asymptotic ALG symmetries.

If the coordinates $x$ and $t$ are periodic, the transverse space to the $w$%
-plane will be $T^{2}=U(1)\times U(1)$. However, again, let us emphasize
that only one of the $U(1)$ isometries of those which would correspond to
the referred $T^2$ is globally defined for the metric (\ref{stew}). In this
case $\tau $ can be interpreted as the modulus of the tori by restricting it
to the fundamental $SL(2,{\bf Z})$ domain. Under this domain we have the
action of the modular transformations $\tau \rightarrow \tau +1$ and $\tau
\rightarrow -1/\tau $. Function $\tau $ will have certain magnetic source
singularities around the $w$-plane. By going around such singularities one
comes up against a jump $\tau \rightarrow \tau +1$, so that the
singularities behave like $2\pi \tau \sim -i\log (w-w_{i})$. By changing
coordinate in the complex plane the metric can be expressed as 
\[
g_{4}=\frac{|dt+\tau dx|^{2}}{\tau _{2}}+\tau _{2}|h(\xi )|^{2}d\xi d%
\overline{\xi }, 
\]%
being $h(\xi )$ an holomorphic function. If we require this metric to be
modular invariant we obtain certain restrictions over $\tau $ and $h(\xi )$ 
\cite{Vafayau}. This requirement implies that $\tau $ is given by the
rational function%
\[
j(\tau )=\frac{P(\xi )}{Q(\xi )}, 
\]%
being $P$ and $Q$ polynomials in the complex variable $\xi $. The function $%
h $ is related to $\tau $ by 
\[
h(w)=\eta (\tau )\overline{\eta }(\tau )\prod_{k=1}^{N}(\xi -\xi
_{i})^{-1/12} 
\]%
where $\eta (\tau )$ is the Dedekind function 
\[
\eta (\tau )=q^{1/24}\prod_{n}(1-q^{n}),\qquad q=\exp (2\pi i\tau ), 
\]%
and $N$ is the number of singularities $\xi _{i}$ of the function $\tau $.
This corresponds to the seven brane solution in \cite{Vafayau}. The
corresponding 4-metric $g_{4}$ is hyper-K\"{a}hler with respect to the
triplet 
\[
\overline{J}_{1}^{\ast }=\Re \left( dt\wedge dx+\frac{1}{2}\tau |h(\xi
)|^{2}d\xi \wedge d\overline{\xi }\right) 
\]%
\begin{equation}
\overline{J}_{2}^{\ast }=\Im \left( \Upsilon \right) ,\qquad \overline{J}%
_{3}^{\ast }=\Re \left( \Upsilon \right) ,  \label{pop}
\end{equation}%
\[
\Upsilon =h(\xi )dt\wedge d\xi +dx\wedge dT. 
\]%
We see that only the first 2-form in (\ref{pop}) is modular invariant,
therefore when modular invariant hyper-K\"{a}hler metrics are extended to a $%
G_{2}$ holonomy one, the modular invariance is generically lost through the
extension.

\section{$G_2$ holonomy metrics fibered over gravitational instantons}

\subsection{The Taub-Nut case}

The family of $G_{2}$ metrics presented in the previous subsection was
constructed with a toric hyper-K\"{a}hler basis whose corresponding Killing
vectors were tri-holomorphic, i.e. they preserved the hyper-K\"{a}hler
triplet (\ref{transurop3}). For this reason, they turned out to be
isometries of the full $G_{2}$ metric. The following examples we consider do
deal with toric hyper-K\"{a}hler metrics, but possessing only one
tri-holomorphic Killing vector. Therefore only this vector will extend the
isometry of the resulting seven-dimensional metric, and the isometry group
will be $U(1)\times {\bf R}^{2}$. As the first example of a less simple
family, let us consider the Gibbons-Hawking metrics (\ref{ashgib}), and
select $V$ as the potential for the electric field of certain configuration
of charges. Then it follows from (\ref{Gibb-Hawk}) that $A$ will be a
Wu-Yang potential describing a configuration of Dirac monopoles located at
the same position where the electric charges are. For a single monopole
located at the origin the potentials will take the form 
\begin{equation}
V=1+\frac{a}{r},\qquad A=\frac{a(ydx-xdy)}{r(r+z)}\qquad z>0,\qquad 
\widetilde{A}=\frac{a(ydx-xdy)}{r(r-z)}\qquad z\leq 0;  \label{scalar}
\end{equation}%
where we have defined here the radius $r^{2}=x^{2}+y^{2}+z^{2}$. The vector
potential is not globally defined in ${\bf R}^{3}$ due to the string
singularities in the $z$ axis. In the overlapping region, the potential $A$
and potential $\widetilde{A}$ differ one to each other by a gauge
transformation of the form $\widetilde{A}=A-2a\ d\arctan (y/x)$. Besides, if
a further gauge transformation $A\rightarrow A+a\ d\arctan (y/x)$ and $%
\widetilde{A}\rightarrow \widetilde{A}-a\ d\arctan (y/x)$ is performed, the
vector potential will be given by a single expression, namely 
\begin{equation}
A=\frac{az}{r}\ d\arctan (y/x);  \label{poto}
\end{equation}%
nevertheless, the potential (\ref{poto}) is clearly discontinuous at the
origin, as it can be seen by evaluating the limit $\delta
A=\lim_{z\rightarrow 0^{+}}A-\lim_{z\rightarrow 0^{-}}A=2a$ $d\arctan (y/x).$
As usual, we are assuming that this limit is taken by crossing the origin
along the $z$ axis. The upper and lower limits are different but related by
a gauge transformation, as expected. The Gibbons-Hawking metric
corresponding to this single monopole configuration is the Taub-Nut
self-dual instanton. The local form of its metric, written in Cartesian
coordinates, reads 
\begin{equation}
g=\left( \frac{r}{r+a}\right) \left( d\tau +\frac{az}{r}d\arctan
(y/x)\right) ^{2}+\left( \frac{r+a}{r}\right) \left(
dx^{2}+dy^{2}+dz^{2}\right) .  \label{Taub-Nut}
\end{equation}%
Although the function $z/r$ is discontinuous, the metric tensors
corresponding to the regions $z>0$ and $z\leq 0$ can be joined to form a
globally defined regular metric by defining a new variable $\widetilde{\tau }%
=\tau +2a\arctan (y/x)$ in the region $z\leq 0$. This implies that the
variable $\tau $ turns out to be periodic. To see this clearly it is
convenient to introduce cylindrical coordinates 
\[
x=\rho \cos \varphi ,\qquad y=\rho \sin \varphi ,\qquad z=\eta ,\qquad
d\arctan (y/x)=d\varphi ; 
\]%
and then it follows that $\widetilde{\tau }=\tau +2a\varphi $. The angle $%
\varphi $ is periodic with period $2\pi $ and therefore the coordinate $\tau 
$ should be actually periodic with period $4a\pi $. Then $\tau $ can be
interpreted as an angular coordinate if the parameter $a$ is an integer
number, $a\in {\bf Z}_{\neq 0}$.

The explicit form for the hyper-K\"{a}hler triplet for the Taub-Nut metric (%
\ref{Taub-Nut}) is given by%
\[
\overline{J}_{1}=\left( d\tau +\frac{az}{r}d\arctan (y/x)\right) \wedge
dx-\left( 1+\frac{a}{r}\;\right) \;dy\wedge dz,
\]%
\begin{equation}
\overline{J}_{2}=\left( d\tau -\frac{az}{r}d\arctan (y/x)\right) \wedge
dy-\left( \;1+\frac{a}{r}\;\right) \;dz\wedge dx,  \label{transurop2}
\end{equation}%
\[
\overline{J}_{3}=\left( d\tau +\frac{az}{r}d\arctan (y/x)\right) \wedge
dz-\left( \;1+\frac{a}{r}\;\right) \;dx\wedge dy.
\]%
Then it is elementary to find the integral forms $H_{i}$. These are 
\[
H_{1}=-x\;d\tau +\left( \;a\ \log (r+z)+z\;\right) dy-a\;x\;d\arctan (y/x),
\]%
\begin{equation}
H_{2}=-y\;d\tau -\left( a\ \log (r+z)+z\right) \;dx-a\;y\;d\arctan (y/x),
\label{integral}
\end{equation}%
\[
H_{3}=-z\ d\tau -a\;r\;d\arctan (y/x)-x\;dy,
\]%
up to a total differential term.

Let us show that the one monopole solution actually corresponds to the
Taub-Nut hyper-K\"{a}hler metric. In spherical coordinates 
\[
x=r\sin \theta \cos \varphi ,\qquad y=r\sin \theta \sin \varphi ,\qquad
z=r\cos \theta ,\qquad d\arctan (y/x)=d\varphi , 
\]%
the metric tensor acquires the form 
\begin{equation}
g=\left( \frac{r}{r+a}\right) \left( d\tau +a\cos \theta d\varphi \right)
^{2}+\left( \frac{r+a}{r}\right) (dr^{2}+r^{2}(d\theta ^{2}+\sin ^{2}\theta
d\varphi ^{2})).  \label{Taub-Nut3}
\end{equation}%
Next, by defining a new radial coordinate $R=2r+a$, it becomes

\begin{equation}
g=\left( \frac{R-a}{R+a}\right) (d\tau +a\cos \theta d\varphi )^{2}+\frac{1}{%
4}\left( \frac{R+a}{R-a}\right) dR^{2}+\frac{1}{4}(R^{2}-a^{2})(d\theta
^{2}+\sin ^{2}\theta d\varphi ^{2}),  \label{Taub-Nut4}
\end{equation}%
which turns out to be the most familiar expression for the Taub-Nut
instanton. Nevertheless, this metric is defined only in the domain $R>a$ and
therefore it has no well defined flat limit $a^{2}\rightarrow \infty $.

With the help of equation (\ref{integral}), it is not difficult to find the $%
G_{2}$ holonomy metrics (\ref{metricol}) fibered over the Taub-Nut
instanton. The first one is thus obtained by selecting $H_{1}^{\ast }=H_{1}$
and $H_{2}^{\ast }=H_{2}$, and the result is 
\[
g_{7}=\mu \;(M+Q\mu )\;\left( \frac{r}{r+a}\right) \;(d\tau
+A(x,y,z))^{2}+\mu \;(M+Q\mu )\;\;\left( \frac{r+a}{r}\right)
\;(\;dx^{2}+dy^{2}+dz^{2}\;)+ 
\]%
\begin{equation}
+\mu ^{-2}\left( \;d\alpha +y\;G(x,y)+F(r,z)\;dx\right) ^{2}+\mu ^{2}(M+Q\mu
)^{2}d\mu ^{2}+  \label{egol}
\end{equation}%
\[
+\frac{1}{\left( M+Q\mu \right) ^{2}}\left( \;d\beta -Qx\;d\tau
+Q\;F(r,z)\;dy-Qa\;x\;d\arctan (y/x)\right) ^{2}, 
\]%
where the function $F$ and the 1-forms $G$ and $A$ are given by%
\[
F(r,z)=a\ \log (r+z)+z,\qquad G(x,y)=d\tau +a\;d\arctan (y/x)\qquad A(x,y,z)=%
\frac{az}{r}\;d\arctan (y/x), 
\]%
and where $Q$ and $M$ are two positive real parameters. Again, for $M>0$, it
can be set to 1 without loss of generality. The case $M=0$, on the other
hand, is the one considered in Ref. \cite{Stele}, cf. Eq. (126) there.
Besides, a second $G_{2}$ holonomy metric is obtained by selecting $%
H_{1}^{\ast }=H_{3}$ and $H_{2}^{\ast }=H_{2}$, leading to the form%
\[
g_{7}=\mu \;(M+Q\mu )\;\left( \frac{r}{r+a}\right) \;(d\tau
+A(x,y,z))^{2}+\mu \;(M+Q\mu )\;\;\left( \frac{r+a}{r}\right)
\;(\;dx^{2}+dy^{2}+dz^{2}\;)+ 
\]%
\begin{equation}
+\mu ^{-2}\;\left( d\alpha +y\;G(x,y)+F(r,z)\;dx\right) ^{2}+\mu ^{2}(M+Q\mu
)^{2}d\mu ^{2}  \label{egol2}
\end{equation}%
\[
+\frac{1}{\left( M+Q\mu \right) ^{2}}\;\left( d\beta -Qz\ d\tau
-Q\;a\;r\;d\arctan (y/x)-Q\;x\;dy\;\right) ^{2}. 
\]%
Notice that the metric holding for the case $H_{1}^{\ast }=H_{2}$ and $%
H_{2}^{\ast }=H_{3}$ just corresponds to a permutation of the coordinates in
(\ref{egol2}) and then gives no new geometry. The Killing vectors
corresponding to both metrics above are $\partial _{\tau }$, $\partial
_{\alpha }$ and $\partial _{\beta }$. We have actually worked out these
metrics in other coordinate systems, but the corresponding expressions are
too cumbersome to write here. The curvature tensor corresponding to both
cases is irreducible and the holonomy is exactly $G_{2}$.

\subsection{The Eguchi-Hanson case}

Now, let us discuss the case of two monopoles on the $z$ axis. Without
losing generality, it can be considered that the monopoles are located in
the positions $(0,0,\pm c)$. The potentials for this configurations are 
\[
V=\frac{1}{r_{+}}+\frac{1}{r_{-}},\qquad A=A_{+}+A_{-}=\left( \frac{z_{+}}{%
r_{+}}+\frac{z_{-}}{r_{-}}\right) d\arctan (y/x),\qquad r_{\pm
}^{2}=x^{2}+y^{2}+(z\pm c)^{2}. 
\]%
This case corresponds to the Eguchi-Hanson instanton, whose metric, in
Cartesian coordinates, reads 
\begin{equation}
g=\left( \frac{1}{r_{+}}+\frac{1}{r_{-}}\right) ^{-1}\left( \;d\tau +\left( 
\frac{z_{+}}{r_{+}}+\frac{z_{-}}{r_{-}}\right) \;d\arctan (y/x)\;\right)
^{2}+\left( \;\frac{1}{r_{+}}+\frac{1}{r_{-}}\;\right)
(\;dx^{2}+dy^{2}+dz^{2}\;),  \label{eguchi}
\end{equation}%
where $z_{\pm }=z\pm c$. In order to recognize the Eguchi-Hanson metric in
its standard form it is convenient to introduce a new parameter $a^{2}=8c,$
and the elliptic coordinates defined by \cite{Prasad} 
\[
x=\frac{r^{2}}{8}\sqrt{1-(a/r)^{4}}\sin \varphi \cos \theta ,\quad y=\frac{%
r^{2}}{8}\sqrt{1-(a/r)^{4}}\sin \varphi \sin \theta ,\quad z=\frac{r^{2}}{8}%
\cos \varphi . 
\]%
In this coordinate system it can be checked that 
\[
r_{\pm }=\frac{r^{2}}{8}\left( 1\pm \left( a/r\right) ^{2}\cos \varphi
\right) ,\qquad z_{\pm }=\frac{r^{2}}{8}\left( \cos \varphi \pm
(a/r)^{2}\right) ,\qquad V=\frac{16}{r^{2}}\left( 1-(a/r)^{4}\cos
^{2}\varphi \right) ^{-1}, 
\]%
\[
A=2\;\left( 1-(a/r)^{4}\cos ^{2}\varphi \right) ^{-1}\;\left(
1-(a/r)^{4}\right) \;\cos \varphi \;d\theta , 
\]%
and, with the help of these expressions, it is found 
\begin{equation}
g=\frac{r^{2}}{4}\left( \;1-(a/r)^{4}\;\right) \;(\;d\theta +\cos \varphi
d\tau \;)^{2}+\left( \;1-(a/r)^{4}\;\right) ^{-1}\;dr^{2}+\frac{r^{2}}{4}%
\;(\;d\varphi ^{2}+\sin ^{2}\varphi d\tau \;)  \label{ego}
\end{equation}%
This is actually a more familiar expression for the Eguchi-Hanson instanton,
indeed. Its isometry group is $U(2)=U(1)\times SU(2)/{\bf Z}_{2}$, the same
as the Taub-Nut one. Actually, the Eguchi-Hanson is a limit form of the
Taub-Nut instanton \cite{Eguchi-Hanson}. The holomorphic Killing vector is $%
\partial _{\tau }$. This space is asymptotically locally Euclidean (ALE),
which means that it asymptotically approaches the Euclidean metric; and
therefore the boundary at infinity is locally $S^{3}$. However, the
situation is rather different in what regards its global properties. This
can be seen by defining the new coordinate 
\[
u^{2}=r^{2}\left( 1-(a/r)^{4}\right) 
\]%
for which the metric is rewritten as%
\begin{equation}
g=\frac{u^{2}}{4}\;(\;d\theta +\cos \varphi d\tau \;)^{2}+\left(
\;1+(a/r)^{4}\;\right) ^{-2}\;du^{2}+\frac{r^{2}}{4}\;(\;d\varphi ^{2}+\sin
^{2}\varphi d\tau \;).  \label{ego2}
\end{equation}%
The apparent singularity at $r=a$ has been moved now to $u=0$. Near the
singularity, the metric looks like 
\[
g\simeq \frac{u^{2}}{4}\;(\;d\theta +\cos \varphi d\tau \;)^{2}+\frac{1}{4}%
du^{2}+\frac{a^{2}}{4}\;(\;d\varphi ^{2}+\sin ^{2}\varphi d\tau \;), 
\]%
and, at fixed $\tau $ and $\varphi ,$ it becomes 
\[
g\simeq \frac{u^{2}}{4}\;d\theta ^{2}+\frac{1}{4}du^{2}. 
\]%
This expression \textquotedblleft locally\textquotedblright\ looks like the
removable singularity of ${\bf R}^{2}$ that appears in polar coordinates.
However, for actual polar coordinates, the range of $\theta $ covers from $0$
to $2\pi $, while in spherical coordinates in ${\bf R}^{3},$ $0\leq \theta
<\pi $. This means that the opposite points on the geometry turn out to be
identified and thus the boundary at infinite is the lens space $S^{3}/{\bf Z}%
_{2}$. For an arbitrary ALE space, the boundary will be $S^{3}/\Gamma ,$
being $\Gamma $ a finite subgroup that induces the identifications. In
general, the multi Taub-Nut metrics, corresponding to $V$ with no constant
term, will be ALE spaces \cite{GibboHa}. In particular, any ALE space admits
an unique self-dual metric \cite{Kronheimer}.

The expressions for the integral forms corresponding to the Eguchi-Hanson
metrics is a bit longer than those of the Taub-Nut case; yielding%
\begin{equation}
H_{1}=-x\;d\tau +\left( \log (r_{+}+z_{+})+\log (r_{-}+z_{-})\right)
dy-2a\;x\;d\arctan (y/x),  \label{ssssrrrr}
\end{equation}%
\begin{equation}
H_{2}=+y\;d\tau \;+\left( \log (r_{+}+z_{+})+\log (r_{-}+z_{-})\right)
\;dx+2a\;y\;d\arctan (y/x),  \label{integrals}
\end{equation}%
\begin{equation}
H_{3}=-zd\tau -a\;(\;r_{+}+r_{-}\;)\;d\arctan (y/x).  \label{rrrrssss}
\end{equation}

By using the formulas (\ref{integrals}) and (\ref{metricol}) the following $%
G_{2}$ holonomy metric is found%
\[
g_{7}=\mu \;(M+Q\mu )\;\left( \;\mu \;(M+Q\mu )\;d\mu ^{2}+\left( \frac{1}{%
r_{+}}+\frac{1}{r_{-}}\right) ^{-1}\left( \;d\tau +\left( \frac{z_{+}}{r_{+}}%
+\frac{z_{-}}{r_{-}}\right) \;d\arctan (y/x)\;\right) ^{2}\;\right) 
\]%
\begin{equation}
+\mu (M+Q\mu )\left( \frac{1}{r_{+}}+\frac{1}{r_{-}}\right)
(\;dx^{2}+dy^{2}+dz^{2}\;)+\frac{(\;d\beta +Q\;H_{1}^{\ast })^{2}}{(\;M+Q\mu
\;)^{2}}+\frac{(\;d\alpha +H_{2}^{\ast }\;)^{2}}{\mu ^{2}}.
\label{metricoli}
\end{equation}%
As before, $H_{1}^{\ast }$ and $H_{2}^{\ast }$ are any pair of 1-forms
selected among those in (\ref{ssssrrrr})-(\ref{rrrrssss}), and there are
essentially two different metrics coming from these possible choices. The
Killing vectors of (\ref{metricoli}) turn out to be the same as those for
the Taub-Nut case; namely, $\partial _{t}$, $\partial _{\alpha }$ and $%
\partial _{\beta }$.

\subsection{Relation to the Ward metrics}

In the previous subsections we considered an array of one and two monopoles
along one axis. In this one we consider an arbitrary array along one axis.
This case also corresponds to axially symmetric hyper-K\"{a}hler metrics,
such as the Eguchi-Hanson and Taub-Nut instantons. Thus, it is convenient to
write the flat three dimensional metric in cylindrical coordinates 
\[
dx^{2}+dy^{2}+dz^{2}=d\rho ^{2}+d\eta ^{2}+\rho ^{2}d\varphi ^{2}. 
\]%
Then the potentials $V$ and $A,$ and consequently the hyper-K\"{a}hler
metrics corresponding to such an array, will be $\varphi $-independent. This
means that $\partial _{\varphi }$ is a Killing vector but, unlike $\partial
_{t}$, it is not tri-holomorphic. For this reason, $\partial _{\varphi }$
will not be necessarily a Killing vector of the full $G_{2}$ metric. In this
sense, there is no advantage in considering these toric 4-metrics if one
wants to enlarge the isometry group of the $7$-metrics. One can consider a
configuration with at least three monopoles that are not aligned \cite%
{Emparan,Gauntlett}, the resulting metrics will not be toric, but the
isometry group of the $G_{2}$ holonomy space will be the same; that is, $%
\partial _{t}$, $\partial _{\alpha }$ and $\partial _{\beta }$.

The interest in considering the toric hyper-K\"{a}hler examples is mainly
that they already encode many well known hyper-K\"{a}hler examples. Such
spaces corresponding to aligned monopoles are known as the Ward spaces \cite%
{Ward}. The solution of the Gibbons-Hawking equation in this case reads $%
A=\rho U_{\rho }d\varphi $ and $V=U_{\eta }$, being $U$ a solution of the
Ward monopole equation $(\rho U_{\rho })_{\rho }+(\rho U_{\eta })_{\eta }=0$%
. The local form for such metrics, when written in cylindrical coordinates,
is%
\begin{equation}
g=\frac{(dt+\rho U_{\rho }d\varphi )^{2}}{U_{\eta }}+U_{\eta }(d\rho
^{2}+d\eta ^{2}+\rho ^{2}d\varphi ^{2}).  \label{Wardfor}
\end{equation}%
For instance, the Taub-Nut metric (\ref{Taub-Nut}) would look like%
\begin{equation}
g=\frac{\sqrt{\rho ^{2}+\eta ^{2}}}{a+\sqrt{\rho ^{2}+\eta ^{2}}}\left( dt+%
\frac{\eta }{\sqrt{\rho ^{2}+\eta ^{2}}}d\varphi \right) ^{2}+\frac{a+\sqrt{%
\rho ^{2}+\eta ^{2}}}{\sqrt{\rho ^{2}+\eta ^{2}}}(d\rho ^{2}+d\eta ^{2}+\rho
^{2}d\varphi ^{2}),  \label{Taub-Nut2}
\end{equation}%
where, for instance, we recognize that the expression (\ref{Taub-Nut2}) is
actually of the Ward form (\ref{Wardfor}) for the function $U=\eta +a\log
(\eta +\sqrt{\eta ^{2}+\rho ^{2}})$, which can be checked to satisfy $(\rho
U_{\rho })_{\rho }+(\rho U_{\eta })_{\eta }=0$. The form (\ref{Wardfor}) is
characteristic of a hyper-K\"{a}hler metric with two commuting isometries,
one of which is self-dual while the other is not. Just for completeness, let
us point out that the integral forms (\ref{integral}) for the Taub-Nut
metric would be expressed in cylindrical coordinates as%
\[
H_{1}=-\rho \cos \varphi \;(d\tau+a\;d\varphi)+U(\sin \varphi d\rho +\rho
\cos \varphi d\varphi ), 
\]%
\begin{equation}
H_{2}=+\rho \sin \varphi \;(d\tau+a\;d\varphi)+U\;(\cos \varphi d\rho -\rho
\sin \varphi d\varphi ),  \label{integral2}
\end{equation}%
\[
H_{3}=-\eta \;d\tau\;-a\;\sqrt{\rho ^{2}+\eta ^{2}} \;d\varphi - \rho \cos
\varphi \;(\sin \varphi d\rho +\rho \cos \varphi d\varphi ). 
\]%
On the other hand, the Eguchi-Hanson solution corresponds to 
\[
U=\log \left( \eta -c+\sqrt{(\eta -c)^{2}+\rho ^{2}}\right) +\log \left(
\eta +c+\sqrt{(\eta +c)^{2}+\rho ^{2}}\right) 
\]%
with $c^{2}>0$. That is, we have two point sources on the $\eta $-axis ($z$%
-axis). Also, the case $c^{2}<0$ corresponds to the potential for an axially
symmetric circle of charge, called Eguchi-Hanson metric of the type I, and
which is always incomplete.

Further, let us consider the fundamental solution of the Ward equation,
namely 
\[
U_{i}=a_{i}\log (\eta -\eta _{i}+\sqrt{(\eta -\eta _{i})^{2}+\rho ^{2}}). 
\]%
This corresponds to a monopole of charge $a_{i}$ located in the position $%
\eta _{i}$. It is not difficult to find the explicit expressions for the
forms $H_{i}$ in analogous way to that in the case analyzed before. If we
consider an array of aligned monopoles, the forms $H_{i}$ will be given by%
\[
H_{1}=-\rho \cos \varphi \;(\;dt+Ad\varphi \;)+U(\sin \varphi d\rho +\rho
\cos \varphi d\varphi _{i})+ b \eta (\sin \varphi d\rho +\rho \cos \varphi
d\varphi ), 
\]%
\begin{equation}
H_{2}=-\rho \sin \varphi \;(\;dt+Ad\varphi \;)-U\;(\cos \varphi d\rho -\rho
\sin \varphi d\varphi )+b\eta (\cos \varphi d\rho -\rho \sin \varphi
d\varphi ),  \label{integral3}
\end{equation}%
\[
H_{3}=-\eta \;d\tau-\;\sum_{i}a_{i}\;\sqrt{\rho ^{2}+(\eta -\eta _{i})^{2}}%
\;d\varphi -b\rho \cos \varphi \;(\sin \varphi d\rho +\rho \cos \varphi
d\varphi ). 
\]%
where now $U$ would be given by $U=\sum_{i}U_{i}$ and $A=\sum_{i}a_{i}$. The
parameter $b$ adds a constant to $V$, for the Eguchi-Hanson metric we have $%
b=0$ and for Taub-Nut $b=1$. The case of infinite array of monopoles
periodically distributed over the axis has been also considered in the
literature; and this was within the context of D-instantons in type IIA
superstring theory \cite{Vafa}. All these examples concern the Ward-type
spaces in four-dimensions. From (\ref{metricol}), one can find the
corresponding $G_{2}$ holonomy metrics, which again take the form 
\[
g_{7}=\frac{\mu \;(M+Q\mu )}{U_{\eta }}\;\left( dt+\rho U_{\rho }d\varphi
\right) ^{2}+\mu \;(M+Q\mu )\;U_{\eta }(d\rho ^{2}+d\eta ^{2}+\rho
^{2}d\varphi ^{2}) 
\]%
\begin{equation}
+\frac{(d\beta +QH_{1}^{\ast })^{2}}{(M+Q\mu )^{2}}+\frac{(d\alpha
+H_{2}^{\ast })^{2}}{\mu ^{2}}+\mu ^{2}\;(M+Q\mu )^{2}\;d\mu ^{2},
\label{hand}
\end{equation}%
being $H_{1}^{\ast }$ and $H_{2}^{\ast }$ any pair of forms selected among
those in the hyper-K\"{a}hler triplet. Again, notice that expression (\ref%
{hand}) gives essentially a pair of $G_{2}$ metrics for a given Ward space (%
\ref{Wardfor}).

\section{Non trivial $T^2$ bundle over hyper-K\"{a}hler}

All the $G_{2}$ metrics considered in the previous sections are solutions of
the Apostolov-Salamon system (\ref{ebol}), (\ref{chon}) and (\ref{compota})
together with the condition $d_{M}^{c}u=0$. As we have seen, the equation (%
\ref{chon}) together with the integrability condition for $H_{1}$ implies
that $\partial \widetilde{J}_{1}/\partial \mu $ should be $\mu $-independent
and closed. We have selected the solution $\widetilde{J}_{1}=(M+Q\mu )%
\overline{J}_{1}$, being $\overline{J}_{1}$ a closed two-form. The resulting
base space was found to be hyper-K\"{a}hler with respect to certain triplet
of 2-forms $\overline{J}_{i}$. Nevertheless, there exist in the literature
examples of hyper-K\"{a}hler structures ($g_{4}$, $J_{i}$, $\overline{J}_{i}$%
) which also admit an strictly almost K\"{a}hler structure ($g_{4}$, $J_{0}$%
, $\overline{J}_{0}$) compatible with the opposite orientation defined by $%
\overline{J}_{1}$, $\overline{J}_{2}$, $\overline{J}_{3}$ \cite{Nurowski}-%
\cite{Armstrong}. This means that $\overline{J}_{i}\wedge \overline{J}_{i}=-%
\overline{J}_{0}\wedge \overline{J}_{0}$. Being \textquotedblleft strictly
almost hyper-K\"{a}hler\textquotedblright\ means that the 2-form $\overline{J%
}_{0}$ is closed though the corresponding almost complex structure $J_{0}$
defined by $\overline{J}_{0}=g_{4}(J_{0}\cdot ,\cdot )$ is not integrable.
Thus $J_{0}$ is not a complex structure. If this is the case, then we can
consider the 2-form 
\begin{equation}
\widetilde{J}_{1}=(M+Q\mu )\overline{J}_{1}+(M^{\prime }+Q^{\prime }\mu )%
\overline{J}_{0},  \label{shak}
\end{equation}%
for which the integrability condition $d^{2}H_{1}=0$ is also satisfied. From
(\ref{compota}) it is obtained an algebraic equation for $u$ with solution 
\begin{equation}
u=\mu \;\left( \;(M+Q\mu )^{2}-(M^{\prime }+Q^{\prime }\mu )^{2}\;\right) ,
\label{shate}
\end{equation}%
being $M^{\prime }$ and $Q^{\prime }$ two additional parameters. This
solution again defines a $G_{2}$ holonomy space, which is in principle
different from the one with $Q^{\prime }=M^{\prime }=0$ of the previous
sections. The task of finding the corresponding $G_{2}$ holonomy metric is a
little bit more complicated because the four dimensional base metric $%
\widetilde{g}_{4}(\mu )$ will not be simply given by a $\mu $-dependent
scaling of the hyper-K\"{a}hler metric $g_{4}$, as it was before. It is
better to illustrate how to construct the $G_{2}$ metric with an example.
Let us consider the distance element 
\begin{equation}
g_{4}=x(dx^{2}+dy^{2}+dz^{2})+\frac{1}{x}(dt+\frac{1}{2}zdy-\frac{1}{2}%
ydz)^{2}.  \label{Nuro}
\end{equation}%
It is not hard to see that such metric tensor is of the Gibbons-Hawking type
(\ref{ashgib}) and is therefore hyper-K\"{a}hler. Let us define the positive
and negative oriented triplets 
\[
\overline{J}_{1}^{\pm }=(dt+\frac{1}{2}zdy)\wedge dz\pm xdx\wedge dy, 
\]%
\begin{equation}
\overline{J}_{2}^{\pm }=(dt+\frac{1}{2}zdy-\frac{1}{2}ydz)\wedge dx\pm
xdy\wedge dz,  \label{nuro3}
\end{equation}%
\[
\overline{J}_{3}^{\pm }=(dt-\frac{1}{2}ydz)\wedge dy\pm xdz\wedge dx. 
\]%
Being \textquotedblleft negative oriented\textquotedblright\ means that the
square of such forms is minus the volume form of $M$. The metric (\ref{Nuro}%
) is hyper-K\"{a}hler with respect to the positive oriented triplet. But it
is easy to see that also $d\overline{J}_{1}^{-}=d\overline{J}_{3}^{-}=0$.
One can also consider any rotated 2-form 
\[
\overline{J}_{\theta }^{-}=\cos \theta \overline{J}_{1}^{-}-\sin \theta 
\overline{J}_{3}^{-} 
\]%
where $\theta $ runs from $0$ to $2\pi $. Such forms will be also closed and
we have a whole circle of negative oriented symplectic forms $\overline{J}%
_{\theta }^{-}$. Nevertheless the almost complex structures $J_{\theta }^{-}$
associated to $\overline{J}_{\theta }^{-}$ are not integrable, that is,
their Nijenhuis tensor is not zero. Thus they are not truly complex
structures. This means that the metric (\ref{Nuro}) admits a circle bundle
of negative oriented almost K\"{a}hler structures which are not K\"{a}hler 
\cite{Nurowski}-\cite{Armstrong}. This is the situation that we were talking
about. The structures of this kind are known as strictly almost K\"{a}hler,
for obvious reasons.

Now, let us select $\theta =0$ for simplicity, and take $\overline{J}%
_{1}^{-} $ as $\overline{J}_{0}$ in (\ref{shak}). Then we have 
\begin{equation}
\widetilde{J}_{1}=(M+Q\mu )\overline{J}_{1}^{+}+(M^{\prime }+Q^{\prime }\mu )%
\overline{J}_{1}^{-}.  \label{sumo}
\end{equation}%
It is convenient to introduce the basis 
\begin{equation}
\widetilde{e}^{1}=\frac{(dt+\frac{1}{2}zdy-\frac{1}{2}ydz)}{\sqrt{x}},\qquad 
\widetilde{e}^{2}=\sqrt{x}dz,\qquad \widetilde{e}^{3}=\sqrt{x}dx,\qquad 
\widetilde{e}^{4}=\sqrt{x}dy,  \label{folo}
\end{equation}%
for the metric (\ref{Nuro}). Then the positive and negative oriented triplet
will be written as 
\begin{equation}
\overline{J}_{1}^{\pm }=\widetilde{e}^{1}\wedge \widetilde{e}^{2}\pm 
\widetilde{e}^{3}\wedge \widetilde{e}^{4},\qquad \overline{J}_{2}^{\pm }=%
\widetilde{e}^{1}\wedge \widetilde{e}^{3}\pm \widetilde{e}^{4}\wedge 
\widetilde{e}^{2},\qquad \overline{J}_{3}^{\pm }=\widetilde{e}^{1}\wedge 
\widetilde{e}^{4}\pm \widetilde{e}^{2}\wedge \widetilde{e}^{3},
\label{posoto2}
\end{equation}%
and, combining (\ref{posoto2}) with (\ref{sumo}), we obtain that 
\begin{equation}
\widetilde{J}_{1}=(\delta _{+}M+\delta _{+}Q\mu )\widetilde{e}^{1}\wedge 
\widetilde{e}^{2}+(\delta _{-}M+\delta _{-}Q\mu )\widetilde{e}^{3}\wedge 
\widetilde{e}^{4}.  \label{sumo2}
\end{equation}%
Here we have denoted $\delta _{\pm }M=M\pm M^{\prime }$ and $\delta _{\pm
}Q=Q\pm Q^{\prime }$. By making the redefinitions 
\[
e^{1}=(\delta _{+}M+\delta _{+}Q\mu )^{1/2}\widetilde{e}^{1},\qquad
e^{2}=(\delta _{+}M+\delta _{+}Q\mu )^{1/2}\widetilde{e}^{2} 
\]%
\begin{equation}
e^{3}=(\delta _{-}M+\delta _{-}Q\mu )^{1/2}\widetilde{e}^{3},\qquad
e^{4}=(\delta _{-}M+\delta _{-}Q\mu )^{1/2}\widetilde{e}^{4},  \label{rodo}
\end{equation}%
we see that the 2-form (\ref{sumo2}) becomes 
\[
\widetilde{J}_{1}=e^{1}\wedge e^{2}+e^{3}\wedge e^{4}. 
\]%
Thus $\widetilde{J}_{1}$ is the K\"{a}hler form of the metric $\widetilde{g}%
_{4}(\mu )=\delta _{ab}e^{a}\otimes e^{b}$. The explicit expression for $%
g_{4}(\mu )$ can be obtained from (\ref{rodo}) and (\ref{folo}), the result
is 
\begin{equation}
g_{4}(\mu )=(\delta _{-}M+\delta _{-}Q\mu )\;x\;(\;dx^{2}+dy^{2}\;)+(\delta
_{+}M+\delta _{+}Q\mu )\;\left( \;xdz^{2}+\frac{1}{x}(\;dt+\frac{1}{2}zdy-%
\frac{1}{2}ydz\;)^{2}\;\right) .  \label{bianca}
\end{equation}%
We immediately observe that $\widetilde{g}_{4}(\mu )$ is not proportional to
the hyper-K\"{a}hler metric (\ref{Nuro}) except if $Q^{\prime }=M^{\prime
}=0 $, which corresponds to the cases considered in the previous sections.

Now, we can construct a $G_{2}$ holonomy metric for (\ref{bianca}) by means
of (\ref{senio}), the result is 
\[
g_{7}=\frac{\left( d\beta +H_{1}\right) ^{2}}{(M+Q\mu )^{2}-(M^{\prime
}+Q^{\prime }\mu )^{2}}+\left( \frac{d\alpha +H_{2}}{\mu }\right) ^{2}+\mu
^{2}\;\left( (M+Q\mu )^{2}-(M^{\prime }+Q^{\prime }\mu )^{2}\right) \;\;d\mu
^{2} 
\]%
\begin{equation}
+\;\mu \;(\delta _{-}M+\delta _{-}Q\mu )\;x\;(\;dx^{2}+dy^{2}\;)+\;\mu
\;(\delta _{+}M+\delta _{+}Q\mu )\;\left( \;xdz^{2}+\frac{1}{x}(\;dt+\frac{1%
}{2}zdy-\frac{1}{2}ydz\;)^{2}\;\right) ,  \label{tunder}
\end{equation}%
where the forms $H_{1}$ and $H_{2}$ are defined by 
\begin{equation}
dH_{1}=\partial _{\mu }\widetilde{J}_{1}=Q\overline{J}_{1}^{+}+Q^{\prime }%
\overline{J}_{1}^{-},\qquad dH_{2}=-\overline{J}_{2}^{+}.  \label{impo}
\end{equation}%
It is really not difficult to obtain the explicit expressions of $H_{i}$
from (\ref{nuro3}). Therefore the expression (\ref{tunder}) is explicit. As
before, more metrics can be obtained by selecting another elements of the
hyper-K\"{a}hler triplet in order to solve (\ref{impo}).

There is another way to check that the expression (\ref{bianca}) for the $%
G_{2}$ metric is correct. Let us consider the calibration form (\ref{dale}).
Then from (\ref{shate}), (\ref{shak}) and (\ref{pip}) we find that 
\begin{equation}
\Phi =\left( (M+Q\mu )\overline{J}_{1}^{+}+(M^{\prime }+Q^{\prime }\mu )%
\overline{J}_{1}^{-}\right) \wedge e^{6}+e^{5}\wedge e^{6}\wedge e^{7}
\label{dale3}
\end{equation}%
\[
+\mu \;\sqrt{(M+Q\mu )^{2}-(M^{\prime }+Q^{\prime }\mu )^{2}}\left( \;%
\overline{J}_{2}\wedge e^{7}+\overline{J}_{3}\wedge e^{5}\;\right) . 
\]%
By using (\ref{posoto2}) together with the redefinitions (\ref{rodo}) it can
be checked again that $\Phi $ takes the octonionic form $\Phi
=c_{abc}e^{a}\wedge e^{b}\wedge e^{c}$. The corresponding $G_{2}$ metric is $%
g_{7}=\delta _{ab}e^{a}\otimes e^{b}$ and after some calculation it is found
the expression (\ref{tunder}), which is what we wanted to show.

Although in principle the metric (\ref{tunder}) contains four parameters ($Q$%
, $M$, $Q^{\prime }$, $M^{\prime }$), only two of them are effective ones.
In fact, by a convenient scaling in (\ref{rodo}) we can select $\delta _{+}M$
and $\delta _{-}M$ equal to one, which means that $M=1$ and $M^{\prime }=0$.
Therefore the $G_{2}$ extension presented in this subsection add two
parameters to the 4-dimensional base space, unlike the extensions considered
in previous sections, which did add only one. Nevertheless here we are
imposing much stronger conditions on the 4-base metric. It should be not
only hyper-K\"{a}hler, but also should possess a bundle of opposite oriented
strictly almost K\"{a}hler structures. Only few of such spaces are known in
the literature, and they are usually too simple. For instance, the 4-metric
that we have presented in this subsection is one of the simplest
Gibbons-Hawking ones, and contains no parameters. The resulting $G_{2}$
metric possesses two effective parameters, the same than the $G_{2}$ metrics
presented in the previous sections. Investigating the existence of less
trivial examples of this kind does deserve attention.

\section{Half-flat associated metrics}

Now, let us revisit the discussion of Ref. \cite{Apostol}-\cite{Santillan}
about the half-flat metrics that are related to the $G_2$ spaces discussed
here. For all the $G_{2}$ holonomy spaces presented so far, we can consider
the six-dimensional hyper-surfaces corresponding to the foliation $\mu
=const $. Then it follows from (\ref{metricol}) that the metrics that have
the form%
\begin{equation}
g_{6}=c_{1}(d\beta +QH_{1})^{2}+c_{2}(d\alpha +H_{2})^{2}+\widetilde{g}_{4},
\label{heteroto}
\end{equation}%
are then defined over such hyper-surfaces. Here, $c_{1}$ and $c_{2}$ are
simply constants. As we will show below, metrics (\ref{heteroto}) are
half-flat spaces \cite{Apostol}. These spaces are of interest in physics,
specially in heterotic string compactifications \cite{Zoupanos}-\cite{Gurro2}

It is not difficult to see that there exists a coordinate system for which
metrics (\ref{metricol}) take the simple form 
\begin{equation}
g_{7}=d\tau ^{2}+g_{6}(\tau ),  \label{to}
\end{equation}%
being $g_{6}(\tau )$ a six-dimensional metric depending on $\tau $ as an
evolution parameter. In fact, by introducing the new variable $\tau ,$
defined by 
\begin{equation}
\mu ^{2}\;(M+Q\mu )^{2}\;d\mu ^{2}=d\tau ^{2},  \label{mu}
\end{equation}%
it can be seen that (\ref{metricol}) takes the desired form. Therefore these 
$G_{2}$ holonomy metrics are a wrapped product $Y=I_{\tau }\times N^{\prime
} $ being $I$ a real interval. The coordinate $\tau $ is just a function of $%
\mu $ and is given by 
\begin{equation}
\tau -\tau _{0}=\int \mu \;(M+Q\mu )\;d\mu =\frac{M\mu ^{2}}{2}+\frac{Q\mu
^{3}}{3}.  \label{inot}
\end{equation}%
An aspect to be emphasized is that $g_{6}(\tau )$ is a half-flat metric on
any hypersurface $Y_{\tau }$ for which $\tau $ takes a constant value.
Indeed, the $G_{2}$ structure (\ref{dale}) can be decomposed as 
\begin{equation}
\Phi =\widehat{J}\wedge d\tau +\widehat{\psi }_{3},  \label{jom}
\end{equation}%
\begin{equation}
\ast \Phi =\widehat{\psi }_{3}^{\prime }\wedge d\tau +\frac{1}{2}\widehat{J}%
\wedge \widehat{J},  \label{joma}
\end{equation}%
where we have defined 
\begin{equation}
\widehat{J}=z^{1/2}\;\overline{J}_{3}+z^{-1/2}(d\beta +A_{1})\wedge (d\alpha
+A_{2}),  \label{jj}
\end{equation}%
\begin{equation}
\widehat{\psi }_{3}=z^{-1/2}\;\widetilde{J}_{1}\wedge (d\alpha +A_{2})+\mu 
\overline{J}_{2}\wedge (d\beta +A_{1}),  \label{daj}
\end{equation}%
\begin{equation}
\widehat{\psi }_{3}^{\prime }=\mu \;z^{-1/2}\overline{J}_{2}\wedge (d\alpha
+A_{2})-\mu ^{2}z^{1/2}\widetilde{J}_{1}\wedge (d\beta +A_{1}),  \label{daj2}
\end{equation}
and $z=\mu ^{2}\;(M+Q\mu )^{2}$. Then the $G_{2}$ holonomy conditions $d\Phi
=d\ast \Phi =0$ for (\ref{dale}) yield 
\[
d\Phi =d\widehat{\psi }_{3}+(d\widehat{J}-\frac{\partial \widehat{\psi }_{3}%
}{\partial \tau })\wedge d\tau =0, 
\]%
\[
d\ast \Phi =\widehat{J}\wedge d\widehat{J}+(d\widehat{\psi }_{3}+\widehat{J}%
\wedge \frac{\partial \widehat{J}}{\partial \tau })\wedge d\tau =0. 
\]%
The last equations are satisfied if and only if 
\begin{equation}
d\widehat{\psi }_{3}=\widehat{J}\wedge d\widehat{J}=0  \label{half-flat}
\end{equation}%
for any fixed value of $\tau $, and 
\begin{equation}
\frac{\partial \widehat{\psi }_{3}}{\partial \tau }=d\widehat{J},\qquad 
\widehat{J}\wedge \frac{\partial \widehat{J}}{\partial \tau }=-d\widehat{%
\psi }_{3}.  \label{ovo}
\end{equation}%
These flow equations were considered by Hitchin in a rather different
context, concerning certain Hamiltonian system whose details are not
important here; see \cite{Hitchin} and \cite{Apostol}. Equations (\ref%
{half-flat}) imply that, for every constant value $\tau ,$ the metric $g_{6}$%
, together with $\widehat{J}$, $\widehat{\psi }_{3}$ and $\widehat{\psi }%
_{3}^{\prime }$, form a half-flat or half-integrable structure \cite{Chiozzi}%
. A constant value for $\tau $ implies a constant value for $\mu ,$ and the
generic form of such half-flat metric corresponds to (\ref{heteroto}). The
reasoning presented above can be also applied to the metric (\ref{tunder})
by defining $\tau $ by 
\[
\tau -\tau _{0}=\int \mu \;\sqrt{(1+Q\mu )^{2}-{Q^{\prime }}^{2}\mu ^{2}}%
\;d\mu . 
\]%
As an example, let us consider the $G_{2}$ metric (\ref{plani}). From (\ref%
{heteroto}), the following half-flat space is obtained 
\begin{equation}
g_{6}=c_{1}(\;d\upsilon -Q\;x\;dz-Q\;y\;dt\;)^{2}+c_{2}(d\chi
-y\;dz-x\;dt)^{2}  \label{planig}
\end{equation}%
\[
+(\;dx^{2}+dy^{2}+dz^{2}+dt^{2}\;). 
\]%
There are no technical difficulties in finding the half-flat metrics
corresponding to each $G_{2}$ holonomy metric described along this work.
Therefore a family of half-flat metrics for the stringy cosmic string, the
Eguchi-Hanson, the Taub-Nut, the almost K\"{a}hler and the Ward cases have
been found through this procedure.

\section{Toric $Spin(7)$ holonomy metrics}

\subsection{$Spin(7)$ metrics that are $T^3$ bundle over hyper-K\"ahler}

Now, we will dedicate the last section to try extend the construction
described here to the case of eight-dimensional spaces with special holonomy
in $Spin(7)$. In reference \cite{Stele} a construction of eight-dimensional $%
Spin(7)$ metrics as $T^{3}$ bundles over hyper-K\"{a}hler metrics was
presented. This construction is actually analogous to (\ref{metricol}) for
the $G_{2}$ holonomy case and leads to the following $Spin(7)$ metric 
\begin{equation}
g_{8}=\frac{(d\alpha +H_{1})^{2}}{\mu ^{2}}+\frac{(d\beta +H_{2})^{2}}{\mu
^{2}}+\frac{(d\gamma +H_{3})^{2}}{\mu ^{2}}+\mu ^{6}\;d\mu ^{2}+\mu
^{3}\;g_{4},  \label{metropol}
\end{equation}%
where, as before, the metric $g_{4}$ is hyper-K\"{a}hler and the 1-forms $%
H_{i}$ are given by $dH_{i}=\overline{J}_{i}$. There is not major
difficulties in proving that the holonomy of this metric is in $Spin(7)$.
Indeed, by defining the tetrad basis 
\[
e_{0}=\mu ^{3}d\mu ,\qquad e_{1}=\frac{d\alpha +H_{1}}{\mu },\qquad e_{2}=%
\frac{d\beta +H_{2}}{\mu }\qquad e_{3}=\frac{d\gamma +H_{3}}{\mu },\qquad
e_{i}=\mu ^{3/2}\overline{e}_{i}, 
\]%
where $\overline{e}_{i}$ is a tetrad for the hyper-K\"{a}hler basis and
where the indices run over $i=1,2,3,4$ and $a=1,2,3.$ It follows that the
4-form defined by the dual octonion constants $c_{abcd}$ 
\[
\Phi _{4}=c_{abcd}e^{a}\wedge e^{b}\wedge e^{c}\wedge e^{d}=e^{0}\wedge
e^{1}\wedge e^{2}\wedge e^{3}+\frac{\mu ^{6}}{6}\overline{J}^{a}\wedge 
\overline{J}^{a}+\mu ^{3}(e^{0}\wedge e^{a}+\frac{\epsilon _{abc}}{2}%
e^{a}\wedge e^{b})\wedge \overline{J}^{a}, 
\]%
turns out to be closed. The presence of such closed form is characteristic
from the reduction from $SO(8)$ from $Spin(7)$.

It is possible to deform this metric in order to get a new one, which will
be again a $T^{3}$ bundle over an hyper-K\"{a}hler base, but now containing
two more effective parameters. The natural deformation {\it ansatz} from (%
\ref{metropol}) would be 
\begin{equation}
g_{8}=\mu ^{2}\;(M_{1}+Q_{1}\mu )^{2}\;(M_{2}+Q_{2}\mu )^{2}\;d\mu ^{2}+\mu
\;(M_{1}+Q_{1}\mu )\;(M_{2}+Q_{2}\mu )\;g_{4}  \label{metropol2}
\end{equation}%
\[
+\frac{(d\alpha +Q_{1}\;H_{1})^{2}}{(M_{1}+Q_{1}\mu )^{2}}+\frac{(d\beta
+Q_{2}\;H_{2})^{2}}{(M_{2}+Q_{2}\mu )^{2}}+\frac{(d\gamma +H_{3})^{2}}{\mu
^{2}}, 
\]%
where $M_{i}$ and $Q_{i}$ are four real parameters. By defining the tetrad
basis 
\[
e_{1}=\frac{d\alpha +Q_{1}\;H_{1}}{M_{1}+Q_{1}\mu },\qquad e_{2}=\frac{%
d\beta +Q_{2}\;H_{2}}{M_{2}+Q_{2}\mu }, 
\]%
\[
e_{3}=\frac{d\gamma +H_{3}}{\mu },\qquad e_{i}=\sqrt{\mu \;(M_{1}+Q_{1}\mu
)\;(M_{2}+Q_{2}\mu )}\;\overline{e}_{i}, 
\]%
it can be checked that also the 4-form 
\[
\Phi _{4}=c_{abcd}e^{a}\wedge e^{b}\wedge e^{c}\wedge e^{d}=e^{0}\wedge
e^{1}\wedge e^{2}\wedge e^{3}+\frac{\mu ^{2}\;(M_{1}+Q_{1}\mu
)^{2}\;(M_{2}+Q_{2}\mu )^{2}}{6}\overline{J}^{a}\wedge \overline{J}^{a} 
\]%
\[
+\mu \;(M_{1}+Q_{1}\mu )\;(M_{2}+Q_{2}\mu )(e^{0}\wedge e^{a}+\frac{\epsilon
_{abc}}{2}e^{a}\wedge e^{b})\wedge \overline{J}^{a}, 
\]%
is closed. Therefore, the deformation (\ref{metropol2}) also defines an $%
Spin(7)$ holonomy metric. Although there are four parameters in the
expression (\ref{metropol2}), only two of them are effective. It is easy to
see that, when $M_{1}$ and $M_{2}$ are nonzero, we can set $M_{1}=M_{2}=1$
by rescaling 
\[
\widetilde{g}_{4}\rightarrow M_{1}\;M_{2}\;\widetilde{g}_{4}\qquad
\Rightarrow \qquad H_{1}\rightarrow M_{1}\;M_{2}\;H_{1},\qquad
H_{2}\rightarrow M_{1}\;M_{2}\;H_{2}\qquad H_{3}\rightarrow
M_{1}\;M_{2}\;H_{3} 
\]%
\[
\alpha \rightarrow M_{1}^{2}\;M_{2}\;\alpha ,\qquad \beta \rightarrow
M_{1}\;M_{2}^{2}\;\beta ,\qquad \gamma \rightarrow M_{1}\;M_{2}\;\gamma
,\qquad Q_{i}\rightarrow \frac{Q_{i}}{M_{i}}. 
\]%
It is actually feasible to extend any of the hyper-K\"{a}hler basis
considered along this work to the case of metrics of $Spin(7)$ holonomy. By
constructions, the resulting metrics will possess four commuting Killing
vectors at least. For instance, for the hyper-K\"{a}hler metrics (\ref{stew}%
) the resulting $Spin(7)$ metrics will be given by 
\[
g_{8}=\frac{\;\left( d\beta -Q_{1}\;\Re (xdt+\frac{1}{2}\overline{T}%
dw)\;\right) ^{2}}{(1+Q_{1}\mu )^{2}}+\frac{\left( \;d\alpha -Q_{2}\Re
(wdt+Tdx)\;\right) ^{2}}{(1+Q_{2}\mu )^{2}}+\frac{\left( \;d\alpha -\Im
(wdt+Tdx)\;\right) ^{2}}{\mu ^{2}} 
\]%
\begin{equation}
+\mu ^{2}\;(1+Q_{1}\mu )^{2}\;(1+Q_{2}\mu )^{2}\;d\mu ^{2}+\mu \;(1+Q_{1}\mu
)\;(1+Q_{2}\mu )\;\left( \frac{|dt+\tau dx|^{2}}{\tau _{2}}+\tau _{2}dwd%
\overline{w}\;\right) ,  \label{metricole}
\end{equation}%
being $\tau $ an holomorphic function on the variable $w$ and $T$ its
primitive. This metric possesses five commuting Killing vectors $\partial
_{\alpha }$, $\partial _{\beta }$, $\partial _{\gamma }$, $\partial _{t}$
and $\partial _{x}$. This procedure can be extended to the Eguchi-Hanson,
Taub-Nut and Ward cases straightforwardly. \newline

\subsection{Non trivial $T^3$ bundle over hyper-K\"ahler}

As a second example, let us comment on a case which is a non trivial example
of $T^{3}$ bundle over hyper-K\"{a}hler. We can generalize our discussion of
the section 5 in order to find metrics that are not of the Gibbons-L\"{u}%
-Pope-Stelle type. As before, let us consider an hyper-K\"{a}hler structure (%
$g_{4}$, $J_{i}$, $\overline{J}_{i}$) which also admits an strictly almost K%
\"{a}hler structure ($g_{4}$, $J_{0}$, $\overline{J}_{0}$) compatible with
the opposite orientation defined by $\overline{J}_{1}$, $\overline{J}_{2}$, $%
\overline{J}_{3}$. Let us note that the 4-form $\Phi _{4}$ of an
eight-dimensional metric $g_{8}=\delta _{ab}e^{a}\otimes e^{b}$ can be
expressed as 
\[
\Phi _{4}=e^{0}\wedge e^{1}\wedge e^{2}\wedge e^{3}+\frac{\widetilde{J}%
^{a}\wedge \widetilde{J}^{a}}{6}+(e^{0}\wedge e^{a}+\frac{\epsilon _{abc}}{2}%
e^{b}\wedge e^{c})\wedge \widetilde{J}^{a} 
\]%
where 
\begin{equation}
\widetilde{J}^{1}=e^{5}\wedge e^{6}+e^{7}\wedge e^{8},\qquad \widetilde{J}%
^{2}=e^{5}\wedge e^{7}+e^{8}\wedge e^{6},\qquad \widetilde{J}%
^{3}=e^{5}\wedge e^{8}+e^{6}\wedge e^{7}.  \label{enough}
\end{equation}%
For the deformed Gibbons-L\"{u}-Pope-Stelle metrics we have that 
\begin{equation}
\widetilde{J}^{i}=\mu \;(M_{1}+Q_{1}\mu )\;(M_{2}+Q_{2}\mu )\overline{J}^{i}.
\label{novo4}
\end{equation}%
being $\overline{J}^{i}$ the hyper-K\"{a}hler triplet of the hyper-K\"{a}%
hler base metric. But if also ($g_{4}$, $J_{0}$, $\overline{J}_{0}$) defines
an strictly almost K\"{a}hler structure then equation (\ref{shak}) suggests
us to modify the definition (\ref{novo4}) and consider 
\[
\widetilde{J}^{1}=\mu \;(M_{2}+Q_{2}\mu )\;\bigg(\;(M_{1}+Q_{1}\mu )\;%
\overline{J}^{1}+(M_{1}^{\prime }+Q_{1}^{\prime }\mu )\;\overline{J}^{0}%
\bigg), 
\]%
\begin{equation}
\widetilde{J}^{2}=\mu \;(M_{2}+Q_{2}\mu )\;\sqrt{(M_{1}+Q_{1}\mu
)^{2}-(M_{1}^{\prime }+Q_{1}^{\prime }\mu )^{2}}\;\overline{J}^{2},
\label{novo}
\end{equation}%
\[
\widetilde{J}^{3}=\mu \;(M_{2}+Q_{2}\mu )\;\sqrt{(M_{1}+Q_{1}\mu
)^{2}-(M_{1}^{\prime }+Q_{1}^{\prime }\mu )^{2}}\;\overline{J}^{3}, 
\]%
where $Q_{1}^{\prime }$ and $M_{1}^{\prime }$ are new parameters. There
always exists an einbein $\overline{e}_{i}$ for which the hyper-K\"{a}hler
triplet is written as 
\[
\overline{J}_{0}=\overline{e}^{1}\wedge \overline{e}^{2}-\overline{e}%
^{3}\wedge \overline{e}^{4},\qquad \overline{J}_{1}=\overline{e}^{1}\wedge 
\overline{e}^{2}+\overline{e}^{3}\wedge \overline{e}^{4}, 
\]%
\begin{equation}
\overline{J}_{2}=\overline{e}^{1}\wedge \overline{e}^{3}+\overline{e}%
^{4}\wedge \overline{e}^{2},\qquad \overline{J}_{3}=\overline{e}^{1}\wedge 
\overline{e}^{4}+\overline{e}^{2}\wedge \overline{e}^{3},  \label{posoto6}
\end{equation}%
Then the natural generalization of (\ref{rodo}) for an einbein which
\textquotedblleft diagonalize\textquotedblright\ the tensors $\widetilde{J}%
^{i}$ is given by 
\[
e^{1}=\mu ^{1/2}\;(M_{2}+Q_{2}\mu )^{1/2}(\delta _{+}M_{1}+\delta
_{+}Q_{1}\mu )^{1/2}\overline{e}^{1}, 
\]%
\[
e^{2}=\mu ^{1/2}\;(M_{2}+Q_{2}\mu )^{1/2}(\delta _{+}M_{1}+\delta
_{+}Q_{1}\mu )^{1/2}\overline{e}^{2} 
\]%
\begin{equation}
e^{3}=\mu ^{1/2}\;(M_{2}+Q_{2}\mu )^{1/2}(\delta _{-}M_{1}+\delta
_{-}Q_{1}\mu )^{1/2}\overline{e}^{3},  \label{rodo7}
\end{equation}%
\[
e^{4}=\mu ^{1/2}\;(M_{2}+Q_{2}\mu )^{1/2}(\delta _{-}M_{1}+\delta
_{-}Q_{1}\mu )^{1/2}\overline{e}^{4}, 
\]%
where $\delta _{\pm }M_{1}=M_{1}\pm M_{1}^{\prime }$ and $\delta _{\pm
}Q_{1}=Q_{1}\pm Q_{1}^{\prime }$. In terms of this basis it is easy to check
by using (\ref{posoto6}) and (\ref{rodo7}) that the expressions (\ref{novo})
for $\widetilde{J}_{i}$ take the diagonal form (\ref{enough}), which is what
we need. Also, by analogy with the cases discussed in the previous sections,
we define the 1-forms 
\[
e_{1}=\frac{d\alpha +H_{1}}{\sqrt{(M_{1}+Q_{1}\mu )^{2}-(M_{1}^{\prime
}+Q_{1}^{\prime }\mu )^{2}}},\qquad e_{2}=\frac{d\beta +H_{2}}{%
M_{2}+Q_{2}\mu } 
\]%
\begin{equation}
e_{3}=\frac{d\gamma +H_{3}}{\mu },\qquad e_{0}=\mu \;(M_{2}+Q_{2}\mu )\;%
\sqrt{(M_{1}+Q_{1}\mu )^{2}-(M_{1}^{\prime }+Q_{1}^{\prime }\mu )^{2}}\;d\mu
,  \label{suc}
\end{equation}%
where the forms $H_{i}$ will be given now by the equations 
\begin{equation}
dH_{1}=Q_{1}\overline{J}_{1}+Q_{1}^{\prime }\overline{J}_{0},\qquad
dH_{2}=Q_{3}\overline{J}_{2},\qquad dH_{3}=\overline{J}_{3}.  \label{virt}
\end{equation}%
The form $\Phi _{4}$ corresponding to (\ref{suc}) is 
\[
\Phi _{4}=d\mu \wedge (d\alpha +H_{1})\wedge (d\beta +H_{2})\wedge (d\gamma
+H_{3})+\frac{\mu ^{2}\;(M_{2}+Q_{2}\mu )^{2}}{6}\bigg(\;f^{2}-f^{\prime
}{}^{2}\;\bigg)\overline{J}^{a}\wedge \overline{J}^{a} 
\]%
\[
+(d\beta +H_{2})\wedge (d\gamma +H_{3})\wedge \bigg(\;f\;\overline{J}%
^{1}+f^{\prime }\;\overline{J}^{0}\;\bigg)+(M_{2}+Q_{2}\mu )(d\gamma
+H_{3})\wedge (d\alpha +H_{1})\wedge \overline{J}^{2} 
\]%
\[
+\mu \;(d\alpha +H_{1})\wedge (d\beta +H_{2})\wedge \overline{J}^{3}+\mu
^{2}\;(M_{2}+Q_{2}\mu )^{2}d\mu \wedge (d\alpha +H_{1})\wedge \bigg(\;f\;%
\overline{J}^{1}+f^{\prime }\;\overline{J}^{0}\;\bigg)
\]%
\[
+\mu \;(M_{2}+Q_{2}\mu )\;(f^{2}-f^{\prime }{}^{2})\;\bigg(\mu \;d\mu \wedge
(d\alpha +H_{2})\wedge \overline{J}^{2}+(M_{2}+Q_{2}\mu )\;d\mu \wedge
(d\gamma +H_{3})\wedge \overline{J}^{3}\bigg)
\]%
where we have defined $f=M_{1}+Q_{1}\mu $ and $f^{\prime }=M_{1}^{\prime
}+Q_{1}^{\prime }\mu $. By virtue of (\ref{virt}) it follows that $d\Phi
_{4}=0$, therefore $\Phi _{4}$ defines an $Spin(7)$ holonomy metric. The
expression for the metric $g_{8}=\delta _{ab}e^{a}\otimes e^{b}$ is given by 
\begin{equation}
g_{8}=\frac{(d\alpha +H_{1})^{2}}{(M_{1}+Q_{1}\mu )^{2}-(M_{1}^{\prime
}+Q_{1}^{\prime }\mu )^{2}}+\bigg(\frac{d\beta +H_{2}}{M_{2}+Q_{2}\mu }\bigg)%
^{2}+\bigg(\frac{d\gamma +H_{3}}{\mu }\bigg)^{2}+\mu \;(M_{2}+Q_{2}\mu
)\;g_{4}(\mu ),  \label{yava}
\end{equation}%
where we have defined the one parameter depending four-dimensional metric 
\[
g_{4}(\mu )=(\delta _{+}M_{1}+\delta _{+}Q_{1}\mu )(\overline{e}^{1}\otimes 
\overline{e}^{1}+\overline{e}^{2}\otimes \overline{e}^{2})+(\delta
_{-}M_{1}+\delta _{-}Q_{1}\mu )(\overline{e}^{3}\otimes \overline{e}^{3}+%
\overline{e}^{4}\otimes \overline{e}^{4}). 
\]%
Therefore, if we deal with an hyper-K\"{a}hler basis which is also strictly
almost K\"{a}hler with a K\"{a}hler form with opposite orientation to the
one defined by the hyper-K\"{a}hler triplet, then the expression (\ref{yava}%
) gives an $Spin(7)$ metric if the equations (\ref{virt}) are satisfied.
This result can be applied for instance to the example (\ref{Nuro}). It is
not difficult to check that the number of effective parameters appearing in
this expression is three, we can select $M_{1}=M_{2}=1$ and $M_{1}^{\prime
}=0$ by an appropriate rescaling of coordinates. Metric (\ref{yava}) is
constructed with an einbein $\overline{e}_{i}$ of an hyper-K\"{a}hler
metric, and is a non trivial example of $T^{3}$ bundle over hyper-K\"{a}hler.

\subsection{Almost $G_2$ holonomy hypersurfaces living inside $Spin(7)$
metrics}

All the $Spin(7)$ metrics obtained in the previous subsections are of the
form 
\[
g_8=d\tau+g_7(\tau), 
\]
being $\tau$ certain coordinate. This means that all these spaces are
foliated by equidistant hypersurfaces and the coordinate $\tau$ is the
distance to a fixed hypersurface $M$. For instance for the metrics (\ref%
{metropol}) the coordinate $\tau$ is defined by 
\[
\tau-\tau_0=\frac{\mu^4}{4}. 
\]
In such cases we have that $e^0=d\tau$ and that the eight-space over which
the metric is defined is decomposed as $M_8=I_{\tau}\times M_7$ being $%
I_{\tau}$ a real interval. The closed 4-form 
\[
\Phi_4= e^0\wedge e^1\wedge e^2\wedge e^3+ \frac{\widetilde{J}^a\wedge%
\widetilde{J}^a }{6}+(e^0\wedge e^a+\frac{\epsilon_{abc}}{2}e^b\wedge
e^c)\wedge \widetilde{J}^a 
\]
can be expressed in this coordinates as 
\[
\Phi_4=d\tau\wedge \omega+\ast_7\omega 
\]
where $\ast_7$ is the Hodge star operation defined on $M_7$. The closure of $%
\Phi_4$ implies that 
\[
d(\ast_7\omega)=0, \ \ \ \partial_{\tau}(\ast_7\omega)=d(\omega). 
\]
These equations were considered in \cite{Hitchin}. The second is known as
the gradient flow equation. The first implies that the 3-form $\omega$ is
co-closed. The seven-spaces with this property are called almost $G_2$
holonomy spaces.

Technically, there are no difficulties to find the almost $G_{2}$ holonomy
metrics for the examples presented in this section. The surfaces $\tau $ is
constant are those for which $\mu $ is constant. The expression for these
almost $G_{2}$ metrics 
\begin{equation}
g_{7}=c_{1}(d\alpha +H_{1})^{2}+c_{2}(d\beta +H_{2})^{2}+c_{3}(d\gamma
+H_{3})^{2}+\widetilde{g}_{4},  \label{almost}
\end{equation}%
with $c_{i}$ being constants, $\widetilde{g}_{4}$ being an hyper-K\"{a}hler
metric and where $H_{i}$ refers to the usual one-forms satisfying $dH_{i}=%
\overline{J}_{i}$. In fact, by making use of (\ref{almost}), any of the
hyper-K\"{a}hler metrics presented along this work can be extended to an
almost $G_{2}$ holonomy metric straightforwardly.%
\[
\]

{\bf Acknowledgement:} We are grateful to Sergey Cherkis and Jos\'{e}
Edelstein for reading the manuscript and for important suggestions; we thank
their interest in our work. We thank the referee of Comm. Math. Phys. for
the corrections and very important remarks. We also thank Jorge Russo for
pointing out interesting references. This work was partially supported by
CONICET and Universidad de Buenos Aires.

\end{document}